\documentclass[11pt,a4paper,british]{article}
\pdfoutput=1
\usepackage{jinstpub}
\usepackage[T1]{fontenc}
\usepackage[utf8]{inputenc}
\usepackage{babel}
\usepackage{float}
\usepackage{textcomp}
\usepackage{amsmath}
\usepackage{amssymb}
\usepackage{upgreek}
\usepackage{graphicx}
\usepackage{eurosym}
\usepackage[per-mode=symbol,separate-uncertainty=true,exponent-product=\cdot,multi-part-units=brackets,product-units=power]{siunitx}
\usepackage[table]{xcolor}
\usepackage{tikz}
\usetikzlibrary{arrows.meta}
\usepackage[super]{nth}
\usepackage[disable]{todonotes} 
\usepackage[skip=1em]{caption}

\usepackage[nottoc,notlof,notlot]{tocbibind}

\usepackage[section,above,below]{placeins}

\usepackage{xspace}
\usepackage{physics}

\title{A compact and light-weight refractive telescope for the observation of extensive air showers}
\author[a]{T.~Bretz}
\author[a]{T.~Hebbeker}
\author[a]{J.~Kemp}
\author[a]{L.~Middendorf}
\author[a]{T.~Niggemann}
\author[a]{C.~Peters}
\author[a,b]{M.~Schaufel}
\author[a]{J.~Schumacher}
\author[b]{J.~Auffenberg} 
\author[b]{C.~Wiebusch}

\affiliation[a]{III. Physikalisches Institut A, RWTH Aachen University, Otto-Blumenthal-Straße, 52074 Aachen, Germany}
\affiliation[b]{III. Physikalisches Institut B, RWTH Aachen University,  Otto-Blumenthal-Straße, 52074 Aachen, Germany}
\emailAdd{tbretz@physik.rwth-aachen.de}

\abstract{A general purpose instrument for imaging of Cherenkov light or fluorescence light emitted by extensive air showers is presented. Its refractive optics allows for a compact and light-weight design with a wide field-of-view of \ang{12}. The optical system features a 0.5\,m diameter Fresnel lens and a camera with 61 pixels composed of Winston cones and large-sized \SI{6x6}{\mm} photo sensors. As photo sensors, semi conductor light sensors (SiPMs) are utilized. The camera provides a high photon detection efficiency together with robust operation. The enclosed optics permit operation in regions of harsh environmental conditions. The low price of the telescope allows the production of a large number of telescopes and the application of the instrument in various projects, such as FAMOUS for the Pierre Auger Observatory, HAWC's Eye for HAWC or IceAct for IceCube.
In this paper the novel design of this telescope and first measurements are presented.

}


\makeatletter
\let\thetitle\@title
\let\themail\@mail
\let\theauthor\@author
\makeatother

\begin{document}
\maketitle

\newpage

\listoftodos

\newpage

\section{Introduction}

One main goal of modern astro-particle physics is the measurement of charged cosmic-rays and photons that reach Earth from outer space. With increasing energy, the flux of these particles is decreasing rapidly 
and direct particle detection would require large detection areas outside of the Earth's atmosphere. Therefore, the indirect detection of cascades from secondary particles that form extensive air showers in the Earth's atmosphere becomes relevant. This technique provides large detection areas that increase with energy, where the measured properties of the extensive air showers reflect the properties of the primary particle.

Particularly interesting is the measurement of optical and near optical light that is emitted from this particle cascades allowing observations far away from the cascades. When relativistic secondary particles of an air shower advance through the Earth's atmosphere, they emit Cherenkov light and trigger the emission of fluorescence light. Experiments detecting this light emission have been carried out for more than 60 years \cite{Lorenz:2012nw,Kampert2012}.

\paragraph*{Fluorescence light} 
The excitation of air molecules by the charged particles of an air shower is followed by the isotropic emission of a few fluorescence photons per meter track length.
Its emission spectrum is dominated by line emission of molecular Nitrogen between 280\,nm and 440\,nm~\cite{Greisen1965}. Its isotropic emission allows the observation of the fluorescence light from all viewing directions but decreases the photon intensity correspondingly.

First attempts by Japanese and US groups to detect fluorescence light from air showers were successful in the early 1970s~\cite{Kampert2012}. This led to large scale experiments such as Fly's Eye~\cite{Baltrusaitis:1985mx} and HiRes~\cite{AbuZayyad:2002sf} and eventually to the fluorescence detectors as part of the Pierre Auger Observatory \cite{Abraham:2009pm} and the Telescope Array \cite{AbuZayyad:2012kk}. This technology shares a significant fraction in the successful measurements of ultra-high energy cosmic rays.

\paragraph*{Cherenkov light} Each relativistic charged particle in the atmosphere with refractive index $n$ and a velocity $v > c/n$ generates Cherenkov photons~\cite{JelleyPorter1963}. A large number of high energy electrons in air showers fulfill this condition. The emitted photons are aligned with the direction of the particle with an opening angle $\theta$ given by $\cos(\theta)=1/(\beta n)$~\cite{Jackson:1998nia}.

For highly relativistic particles ($\beta\to 1$) and a refractive index of air at sea level (n\,$=$\,1.0003) the opening angle reaches its maximum of $\theta\sim$\,\ang{1.4}. The wavelength-spectrum of air-Cherenkov photons is continuously decreasing with $\dd n/\dd\lambda \propto \lambda^{-2}$ \cite{Jackson:1998nia}. This yields an emission of roughly a photon per centimeter track length. The absorption of UV light through air molecules leads to a lower cut-off in the near-UV between \SIrange{300}{350}{\nm} \cite{Baltrusaitis:1985mx}.
Unlike the largely isotropic emission of fluorescence light, the emission into a small solid angle of less than \SI{\sim e-3}{sr} results into a higher photon intensity, if the observer is located within the Cherenkov cone. Therefore, a significantly smaller detection threshold can be achieved by viewing into the direction of the shower. The solid angle from which the emission can be detected is reduced respectively.

Cherenkov light from air-showers is used to image these showers using optical telescopes known as imaging air-Cherenkov telescopes.

With the first detection of a gamma-ray source at TeV energies through the emission of Cherenkov light, the Whipple 10\,m telescope~\cite{Weekes:1989tc} has opened a whole new field, eventually leading to todays high precision instruments H.E.S.S.~\cite{Aharonian:2006pe}, MAGIC~\cite{Albert:2007xh} and VERITAS~\cite{Weekes:2001pd}. As of today, in total more than two hundred sources have been detected to emit very high energy gamma-rays~\cite{TeVcat}. In October 2011, the First G-APD Cherenkov Telescope (FACT,~\cite{FACT}) utilized for the first time a camera based on semi-conductor photo sensors (SiPMs) and benefited from it. The ongoing construction of the Cherenkov Telescope Array (CTA,~\cite{CTA}) comprising telescopes with SiPMs underlines the success of the imaging air-Cherenkov technique in general and the successful application of SiPMs in particular. 

\paragraph{Comparison} 
In both cases, the light emission is dominated by the high energy electrons of the shower. This number scales almost linearly with the energy of the primary particles, and it also depends on the type of the primary particle, the altitude of the shower maximum and other atmospheric and operational conditions. Maximizing the light collection efficiency minimizes the energy threshold if done in a way that the signal-to-noise ratio decreases. A limiting factor is the level of diffuse ambient light from the night-sky. Consequently, both detection methods favor cameras sensitive to single photons. One way to suppress the ambient light is to minimize the possible integration times of the signal.

The observation of fluorescence light from the side allows for a larger field-of-view than the observation of Cherenkov light where the limited opening angle naturally limits the visible image to a few degrees. The optimization of the optics suited for both measurements needs to find a compromise for these different requirements. While for fluorescence light a field-of-view of typically \ang{15} or more is required, the imaging air-Cherenkov technique is satisfied with around \ang{4}. Images of fluorescence showers typically extend over more than \ang{10}, images in imaging air-Cherenkov telescopes have extensions in the order of a degree. The latter renders pixel sizes of sub-degree necessary if the image itself is required to determine the properties of the primary particle and for background suppression. If, for example, the combination with an extensive air-shower array eliminates the need for a precise description of the image, the pixel size can be larger. The same applies for showers at PeV energies where image extensions easily exceed one degree. The presented telescope is a compromise in terms of optics between both techniques suited for the application at TeV energies in coincidence with an extensive air-shower array or as fluorescence telescope.

A similar compromise has to be found in terms of data acquisition for similar geometrical reasons. The minimum required sampling frequency is only determined by the duration of the pulse produced by the light-sensor. In case of a long input signal, a lower sampling frequency can be chosen. While fluorescence telescopes usually work with sampling frequencies of a few hundred mega-samples per second, imaging air-Cherenkov telescope require GHz sampling frequencies to sample the exceptionally short flashes of air-showers. In general, sampling at GHz frequencies works for both techniques. The Cherenkov photons, directed towards an observer, arrive within a few nanoseconds. The arrival times of fluorescence photons can last for a micro-second or more, when the shower is viewed from the side. Consequently, for fluorescence showers, the accessible sampling depth must be significantly larger. These considerations also influence the ideal trigger logic for both cases. A sum-trigger with a small patch size should provide an excellent compromise for both cases.

\section{The telescope}
\begin{figure}[htp]

\begin{minipage}[c]{.5\textwidth}
	\includegraphics[width=\textwidth]{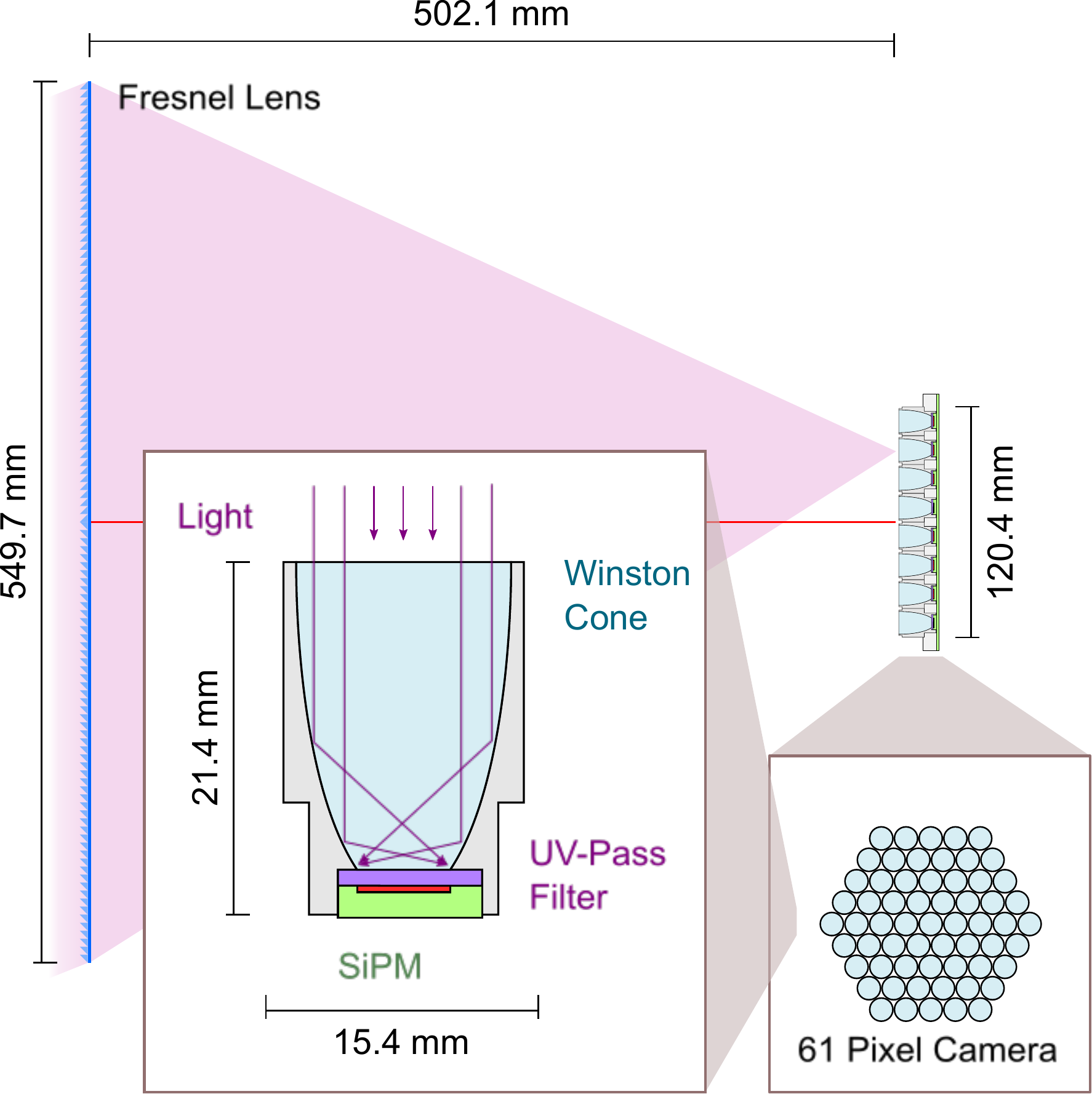}
\end{minipage} 
\hfill 
\begin{minipage}[c]{.5\textwidth}
	\includegraphics[width=\textwidth]{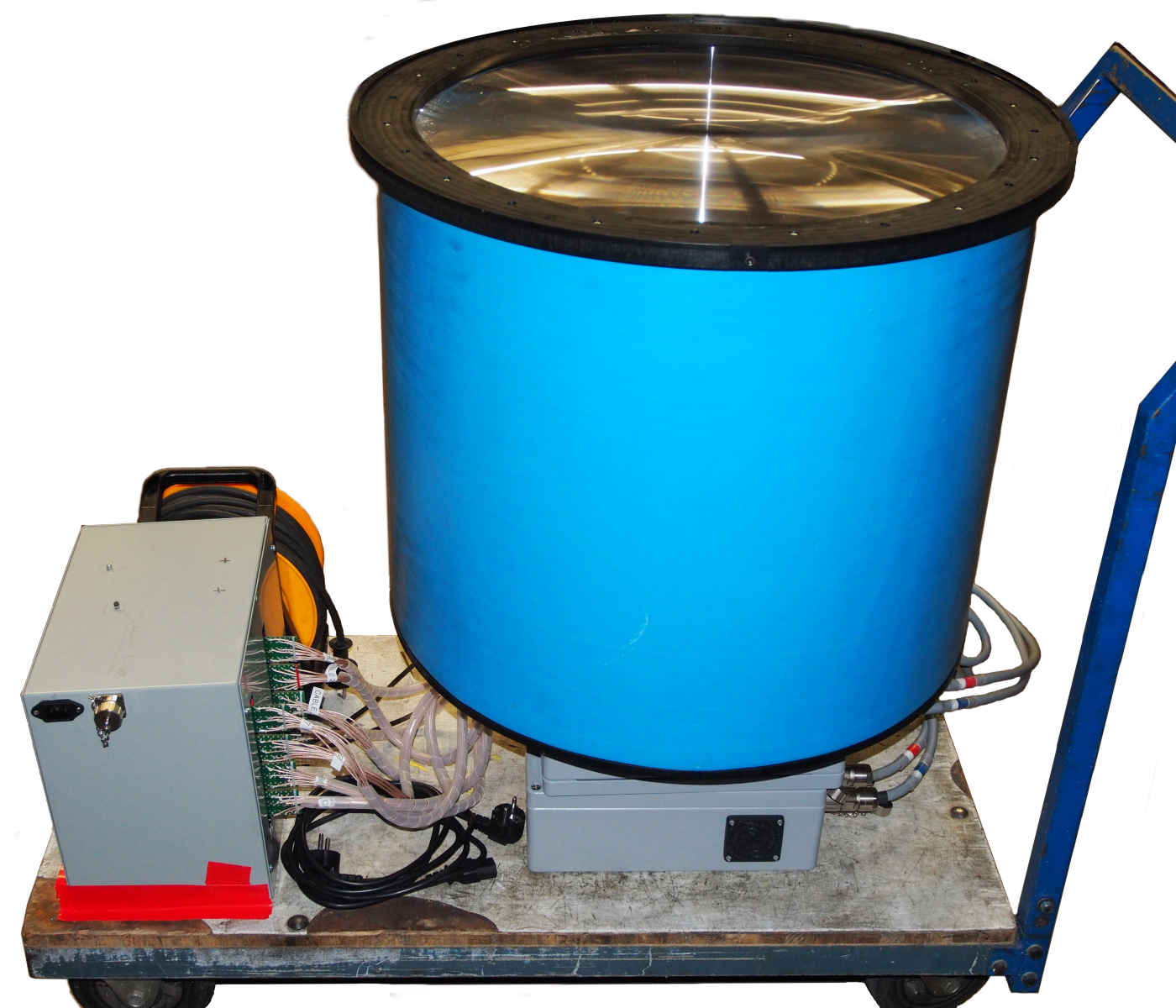}
\end{minipage}
\caption{\label{fig:telescope} \emph{Left:} Schematic overview of the optics of the telescope \cite[c.f.][]{Niggemann2016}. \emph{Right:} A picture of the compact telescope setup during a measurement campaign in front of RWTH Aachen University. The box on the left contains the data acquisition electronics, the box below the tube contains the camera and the power supply.}
\end{figure}

The presented telescope design implements a large field-of-view optimized for fluorescence light detection and a data acquisition with GHz sampling speed optimized for Cherenkov light detection. The concept is sketched in Fig.~\ref{fig:telescope}. The refractive optics is based on a Fresnel lens with an aperture of \SI{0.24}{\square\meter}. The camera has 61 pixels with a field-of-view of about \ang{1.5} each, resulting in a total field-of-view of \ang{12}. 
To lower the energy threshold for the detection of air showers, a future increase of the aperture at still moderate cost is not prohibited by the current design.

Each pixel uses a non-imaging light-guide to increase the sensitive area of each photo sensor. Semi-conductor photo sensors (SiPM) are preferred to classical photo-multiplier tubes (PMT). Today's UV-optimized SiPMs with a p-on-n silicon process offer photo detection efficiencies from \SI{250}{nm} to almost \SI{1000}{nm}~\cite[e.g.][]{SenslJ}. They are tolerant against illumination with bright light allowing for a high duty cycle even under bright light conditions such as moonlit nights~\cite{FACTHighlights}. This leads to an instrument suitable for simple handling and robust operation. The application of typical voltages between \SIrange{30}{100}{\volt} as compared to kilo-volts when using classical photo multipliers, simplifies the operation and reduces costs.

The combination of the Fresnel lens and SiPMs allows a small, light-weight, and enclosed design. This reduces costs but also protects the camera which increases the robustness allowing operation under extreme environmental conditions. Due to its light-weight design (\SI{< 32}{\kg} including mechanics, optics, and electronics) it can be transported and installed easily, enhancing the operational and logistical flexibility. The total price of a fully operational instrument including mechanics, camera, data acquisition and slow control is only at the level of \euro\,\num{10000} (excl.\ VAT) for the current prototypes. Prices are reduced further, once larger quantities are produced or less expensive components are selected.

The instrument was initially developed as a demonstrator for the application of semi-conductor photo sensors (SiPM) in fluorescence telescopes in the context of AugerNext~\cite{Niggemann2013,Haungs2015} and became known as the FAMOUS\footnote{{\bf F}irst {\bf A}uger {\bf M}ulti-pixel photon counter camera for the {\bf O}bservation of {\bf U}ltra-high energy air {\bf S}howers} telescope~\cite{Niggemann2016,Bretz2015}. During the development, it became apparent that the telescope serves equally well, if not better, as a detector for air-Cherenkov light. Due to its small, low-cost and robust design, the telescope is suitable for a large number of applications. The application of hybrid detection of ultra-high energy air-showers allows for the cross-calibration of the energy scale of surfaces detectors in the Pierre Auger Observatory~\cite{PierreAuger}. As \emph{HAWC's~Eye}~\cite{HAWCsEye} it has been proposed as the first hybrid detector for Cherenkov light extending the High Altitude Water-Cherenkov Observatory (HAWC,~\cite{HAWCCrab}) to improve its absolute energy calibration and resolution~\cite{HAWCsEye}. The system has also been proposed as the IceAct~\cite{Auffenberg2017} surface extension of IceCube~\cite{Aartsen:2016nxy}. Here, the goal is to calibrate the directional resolution and detection efficiency of the IceCube and IceTop detectors, to improve cosmic ray composition measurements~\cite{Auffenberg2017b} and to provide cosmic ray vetoing for neutrino observations~\cite{Auffenberg2015}. As another important application, its compactness, price, and easy and safe handling make this instrument ideal for the use in lab courses to educate students in astro-particle physics or as an outreach instrument. Such a lab course is currently under development for our university.

The following chapters will discuss the mechanics, the optics, the data acquisition and slow control electronics, and the software of a first operating demonstrator instrument in details. Finally, first successful measurements in Aachen (Germany) are presented as a proof of concept.

\section{Mechanical Structure}

\begin{figure}[t]
    \includegraphics[height=8cm]{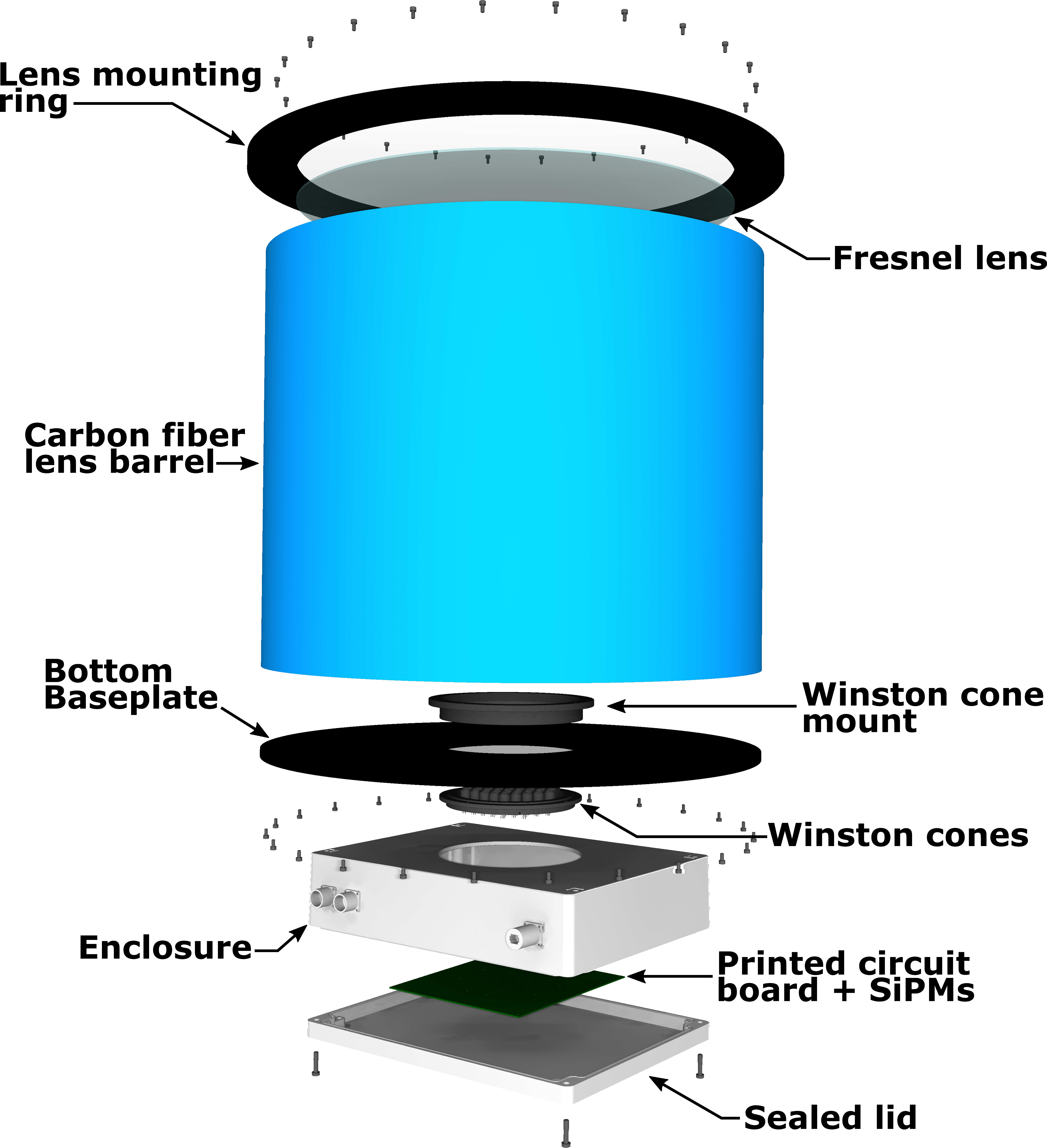}
	\hfill 
	\includegraphics[height=7.5cm]{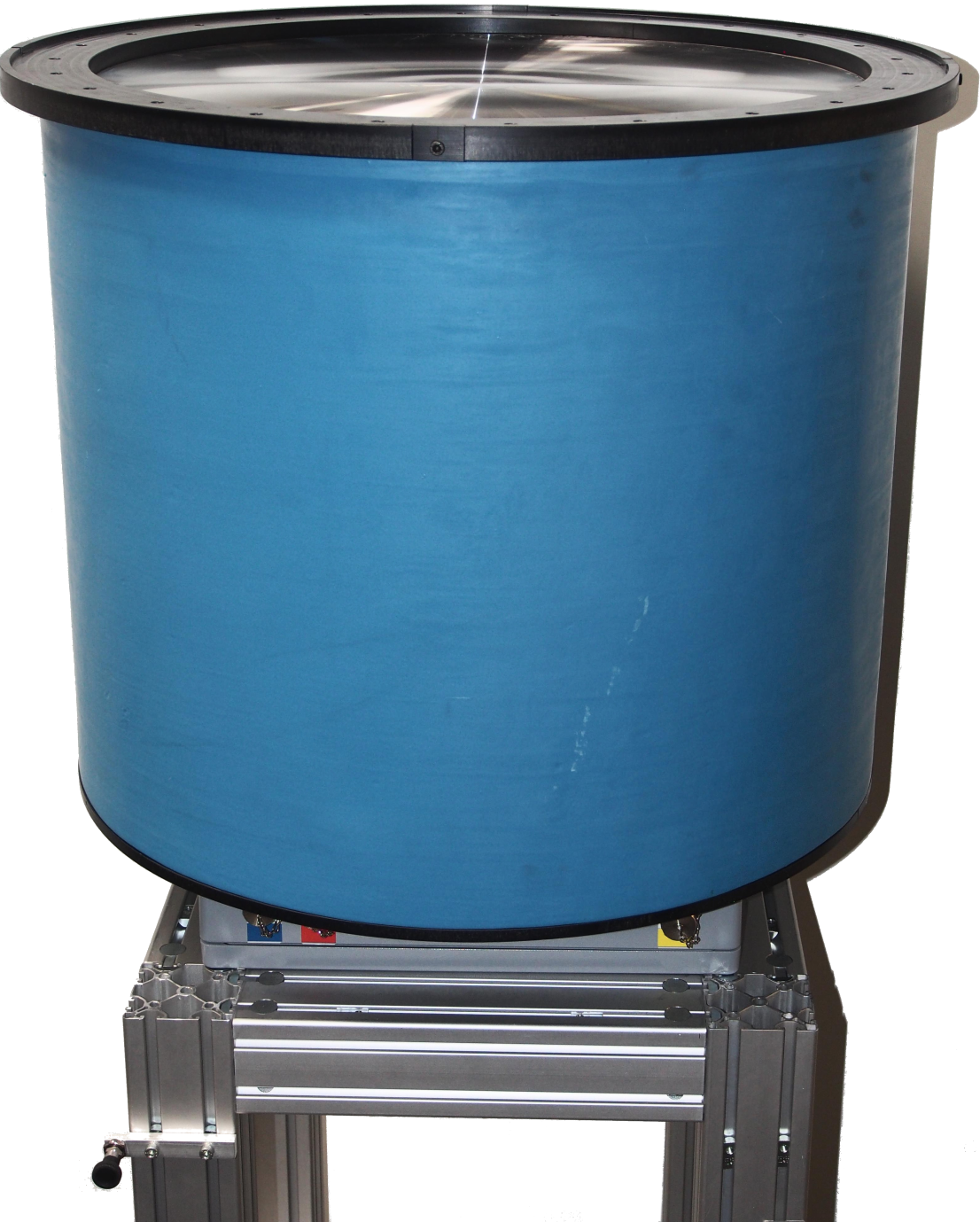}
\caption{\label{fig:telescope2} \emph{Left:} Sketch of the mechanical assembly of the telescope. \emph{Right:} a picture of the fully-assembled telescope.}
\end{figure}

The main elements of the telescope system are the lens, the tube barrel, the focal plane with the camera and the enclosure behind the camera housing the electronics. An overview of the system is shown in Fig.~\ref{fig:telescope2}.

The cylindric tube of the telescope has a diameter of \SI{568}{\mm}, a length of \SI{498}{\mm} and \SI{6}{\mm} wall thickness. It is made out of light-weight carbon compound material to reduce weight (\SI{15}{\kg}) and cost (\euro\,200). One side holds the Fresnel lens where a plastic clamping ring is used to fix the lens. Black velour adhesive foil is used as diffusive inlay on the inner surfaces of the tube to reduce the sensitivity to stray light. A lid can be used to cover the aperture and thus protect the lens from weather conditions and shield the SiPMs on the focal plane from external light. The focal plane is mounted at a distance of one focal length from the lens on the other end of the tube. An  enclosure is mounted behind the focal plane to provide protection of the focal plane against environmental water and to support the read-out electronics.

The choice of mount depends on the application. If star tracking is required, its low weight allows to utilize a regular commercial telescope mount. 

\section{Optics}


In the following, the optical components of the system are described: lens and camera with light concentrators, filters and light sensors. The optical sensitivity curves of these components are summarized in Fig.~\ref{fig:efficiencies}.
A full GEANT\,4-based~\cite{GEANT4} description of the optical system is developed~\cite{Niggemann2013} and is used for simulation studies.

\subsection{Fresnel lens}

\begin{figure}[t]
	\includegraphics[width=.495\textwidth]{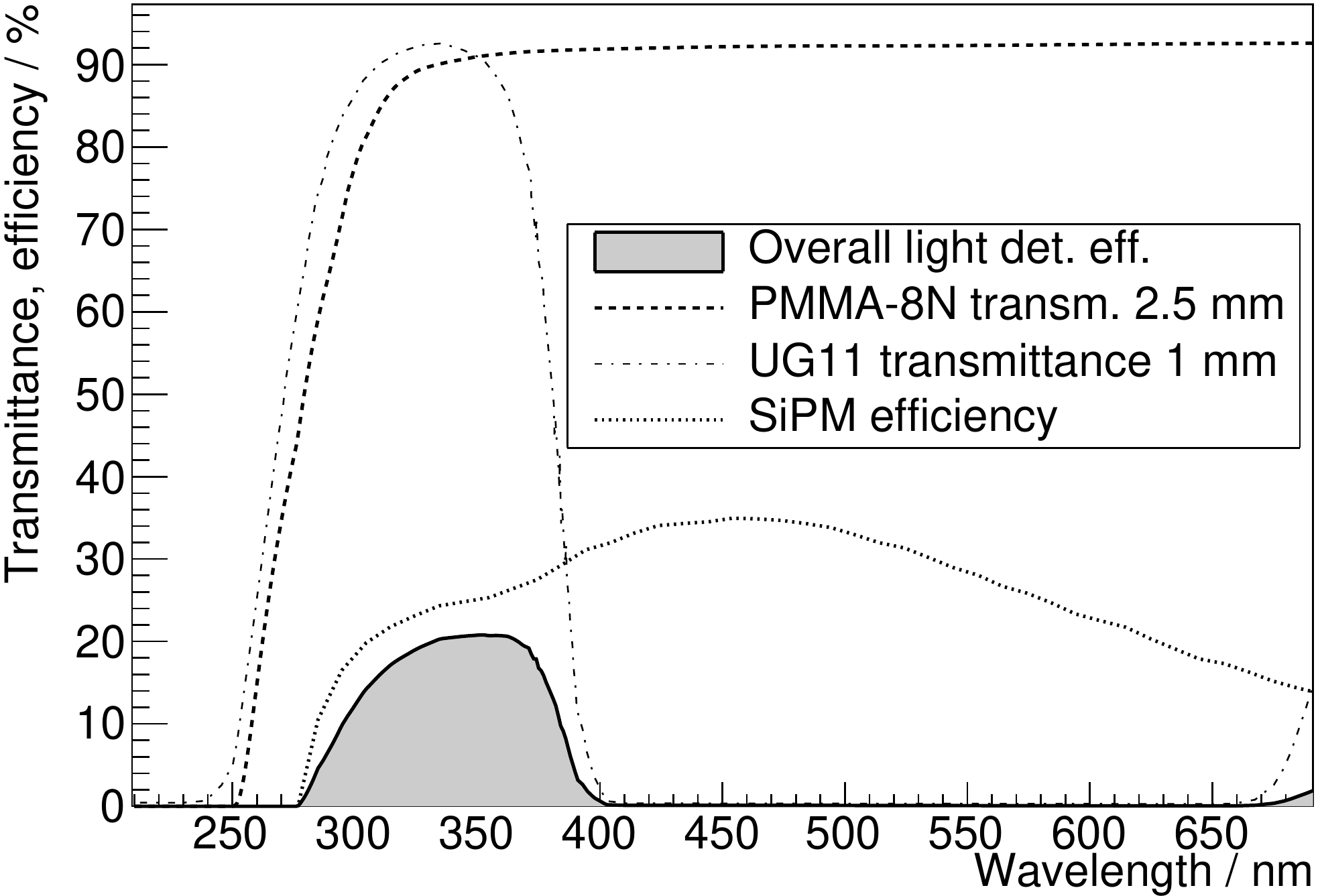}
	\hfill 
	\includegraphics[width=.495\textwidth]{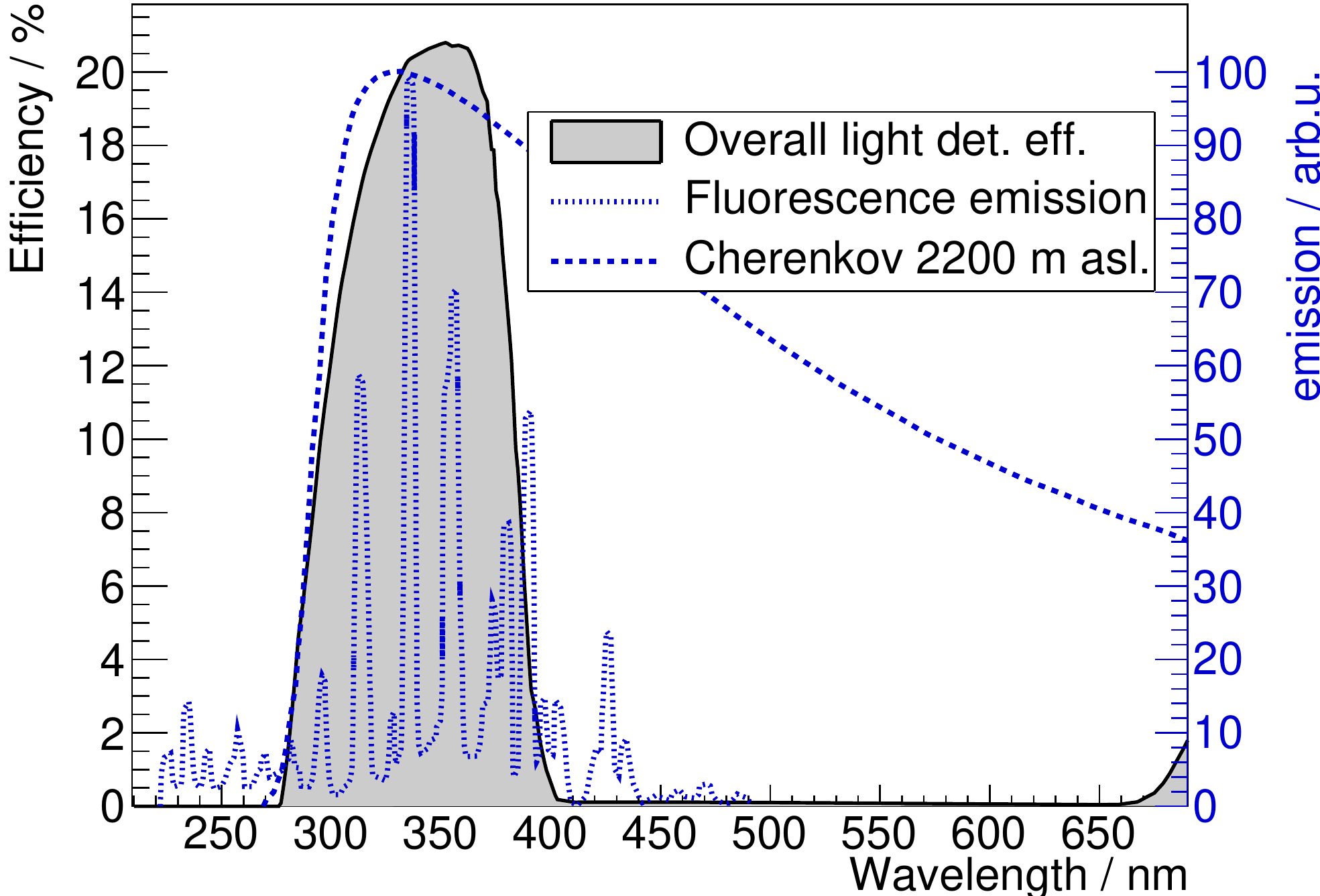}
\caption{\label{fig:efficiencies} Telescope sensitivity curves; \emph{Left:} Transmittance of \SI{2.5}{\mm} Polymethyl\- metha\-crylate (PMMA) including Fresnel reflections -- the base material of the Fresnel lens (dashed) \cite{PMMA}, SiPM photo detection efficiency (dotted) \cite{HamamatsuS10943} and transmittance of UG\,11 UV bandpass filter glasses (dashed-dotted) \cite{UG11} together with the folded combination of all three components. \emph{Right:} Folded combination spectrum together with the spectra of air-Cherenkov light (dashed, at \SI{2200}{\meter} above sea level with O$_2$ and O$_3$ absorption)~\cite{Bouvier2013} and fluorescence emission (dotted)~\cite{Bunner1967}.}
\end{figure}

The light-focusing element of the telescope is a Fresnel lens made of UV-transparent Polymethyl methacrylate (PMMA) and is manufactured by ORAFOL Fresnel Optics GmbH \cite{Orafol}. The small thickness (2.5\,mm) of the lens due to the Fresnel technique significantly reduces its weight and transmission loss compared to plano-convex lenses.
Fig.~\ref{fig:efficiencies} shows the transmittance versus wavelength of the Fresnel lens material. The transmittance of the PMMA used here (PMMA-8N, refractive index 1.53 at 300\,nm and relative dispersion 58) exceeds 90\% in the near UV above 300\,nm including surface reflections and is thus ideally suited for the detection of atmospheric Cherenkov and fluorescence light. The current geometrical implementation of the grooves of the Fresnel lens leads to a total transmission of about \SI{50}{\percent} \cite{EichlerMaster} for vertically incident light.
 This is due to a commercial design not optimized for the current use case and can be improved by a dedicated design ~\cite{OrafolOpt}. 



\begin{figure}[t]
\centering
	\includegraphics[width=.7\textwidth]{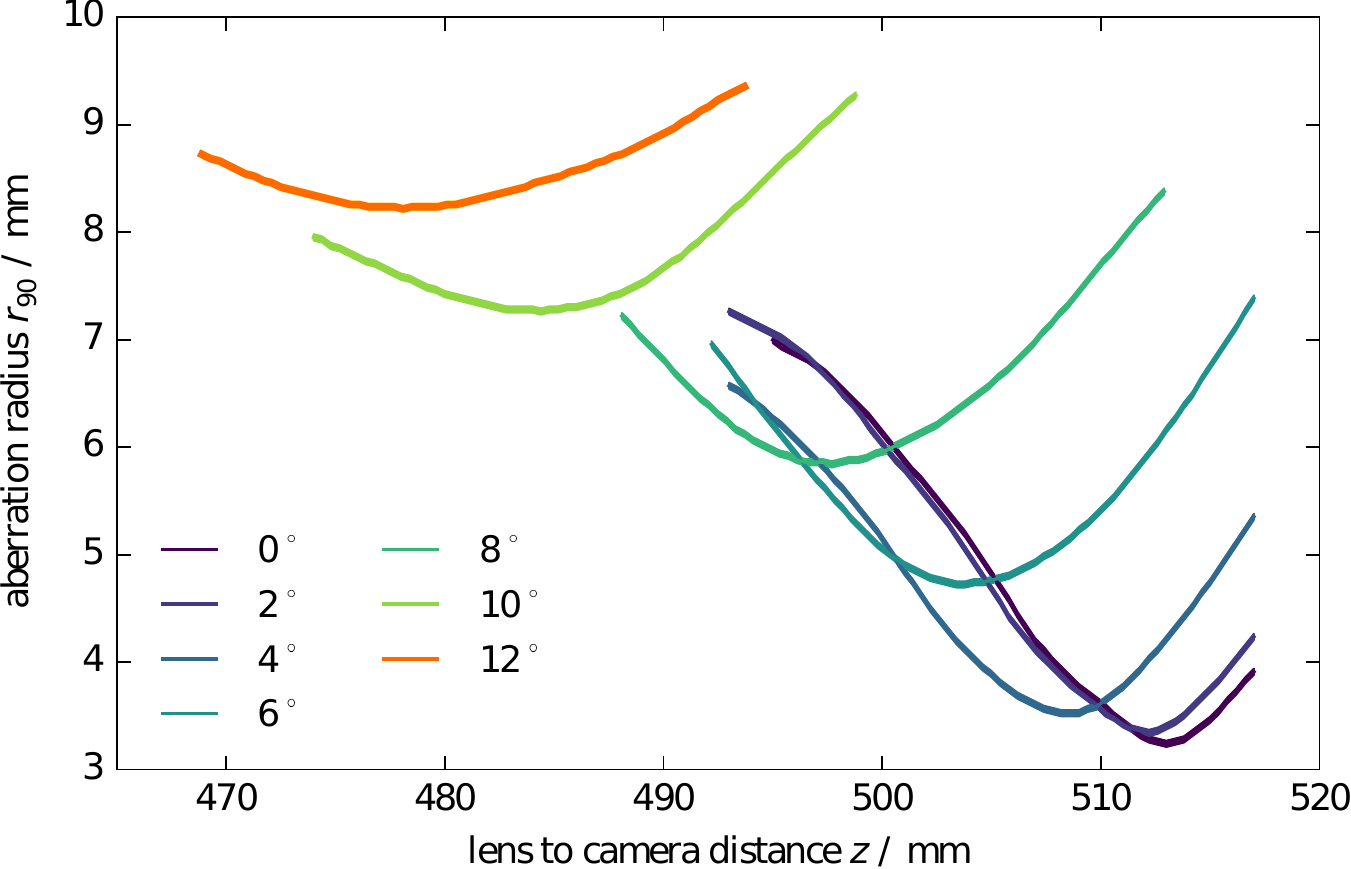}
\caption{\label{fig:lens}
A measurement of the aberration radius vs.\ lens-to-camera distance for incident angles from \ang{0} up to \ang{+-12}. The telescope is optimized for incident angles up to \ang{+-6} yielding a point-spread-function smaller than the radius of one pixel, i.e.\ 6.71\,mm, for the nominal focal length of 502.1\,mm. Measured at a wavelength of \SI{550}{\nm}. Taken from \cite{Niggemann2016}.}
\end{figure}

The diameter of the light-focusing area is 549.7\,mm. According to the manufacturer, the focal length is 502.1\,mm for perpendicular photons at (546$\pm$27)\,nm. This results in $F/D\sim1.1$ which is ideal for wide field-of-view telescopes as image distortion significantly increases for larger apertures ~\cite{BretzRibordy}. Ten grooves per millimeter cover the surface of the lens providing the required imaging quality of \SI{90}{\percent} light contained in one pixel. The point-spread-function was measured for incident angles of up to \ang{+-12}. Fig.~\ref{fig:lens} shows a measurement of the derived aberration radius $r_{90}$ which holds \SI{90}{\percent} of the light intensity as a function of the lens-to-camera distance for different incident angles. For a distance of approximately 508\,mm incident angles up to the nominal field-of-view of \ang{+-6}, the aberration radius is smaller than the radius (\SI{6.7}{\mm}) of the light concentrator of each pixel. This defines the nominal camera-to-lens distance for the design of the optical system. This measurement is in good agreement with values obtained by simulations, see~\cite{Niggemann2016}.

\subsection{The Camera}
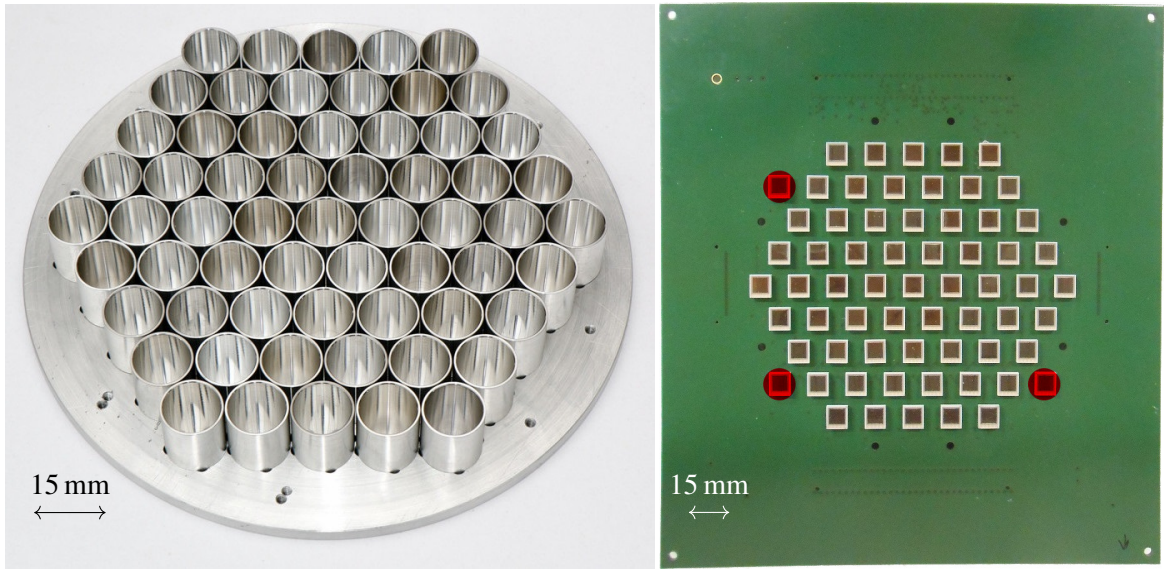
\begin{figure}[thbp]
	  	\begin{tikzpicture}
	    \node[anchor=south west,inner sep=0] (image) at (0,0) {\includegraphics[height=7.48cm]{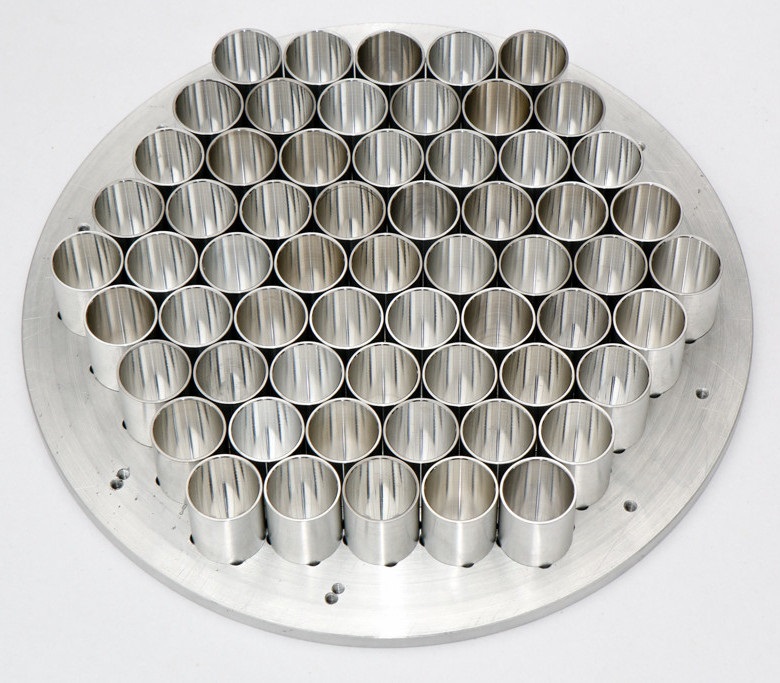}};
	    \begin{scope}[x={(image.south east)},y={(image.north west)}]
        	\draw[<->] (0.1-0.055, 0.1) -- (0.1+0.055, 0.1);
			\node at (0.1,0.1+0.05) {\SI{15}{\milli\meter}};
	    \end{scope}
	\end{tikzpicture}
    \hfill 
    \begin{tikzpicture}
	    \node[anchor=south west,inner sep=0] (image) at (0,0) {\includegraphics[height=7.5cm]{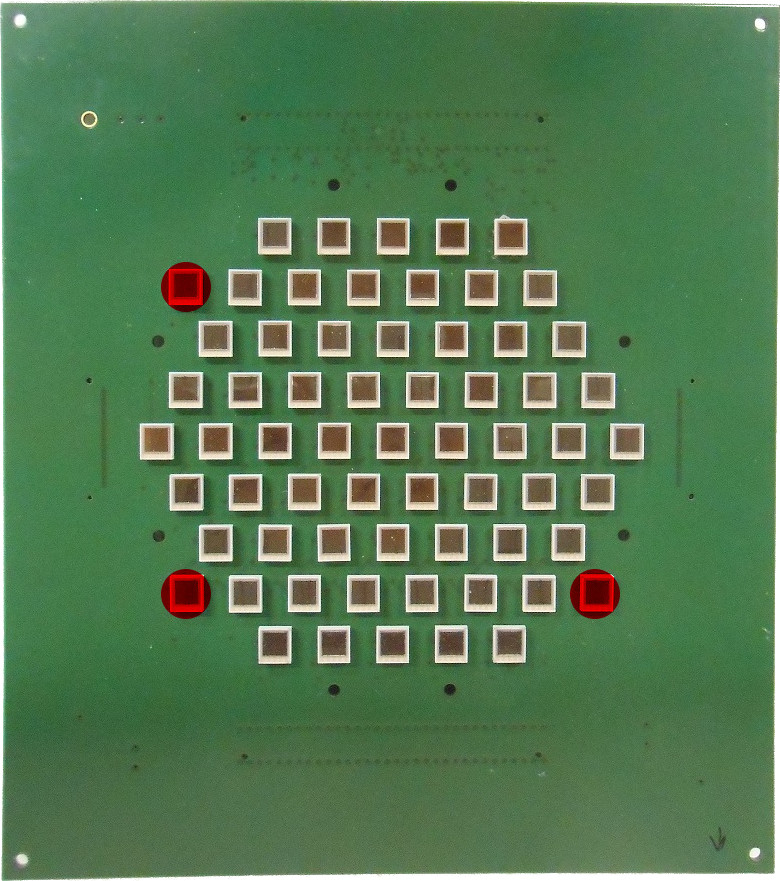}};
	    \begin{scope}[x={(image.south east)},y={(image.north west)}]
        	\draw[<->, color=white] (0.1-0.04, 0.1) -- (0.1+0.04, 0.1);
			\node[color=white] at (0.1,0.1+0.05) {\SI{15}{\milli\meter}};
	    \end{scope}
	\end{tikzpicture}
\caption{\label{fig:focalplane} Aluminum supporting frame for the camera with glued-in Winston cones seen from the side facing the lens (left). The Winston cones focus the light on the photo sensors. The 61 photo sensors mounted on a printed circuit board are shown on the right. The three \textit{blind pixels} are marked in red.}
\end{figure}

The camera (see Fig.~\ref{fig:focalplane}) consists of 61 active pixels. Each pixel uses one \SI{6x6}{\mm} semi-conductor
photo sensor. All photo sensors are mounted on a single printed circuit board (PCB). A light concentrator is mounted in front of each sensor to increase the light collection area.

\subsubsection{Light concentrators and filters}

To increase the sensitive area of the photo sensors, and to match them to the size of the aberration spot, light concentrators are used. These are hollow Winston Cones~\cite{Winston2005} with entrance and exit radii of \SIlist{6.71;3.00}{\mm}, respectively. This leads to a maximum acceptance angle of \ang{26.6}. 

The rotational symmetric design of a Winston Cone was chosen for reasons of simplicity in the initial production of the demonstrator. Follow-up studies~\cite{JanPaulMaster} have shown that a substantial improvement of a factor 1.6 of the light collection can be achieved with hexagonal-to-quadratic solid light concentrators keeping the remaining geometry identical. This optimizes the fill factor for the hexagonal camera layout and reflects the quadratic photo sensitive area. A comprehensive review on a further optimization of an optical system including light-collectors can be found in~\cite{BretzRibordy}.

All cones are made from aluminum that has been milled and then polished on the optical side to perform well in the near ultra-violet range with an efficiency of more than \SI{90}{\percent} (ideal aluminum)~\cite{Rakic1998}. The wall thickness of the aluminum at the entrance side is 0.69\,mm. The minimum wall-to-wall distance between neighboring cones is 0.20\,mm increasing the dead space between the pixels but reducing complexity of the installation of the camera.

The Winston Cones are glued to an aluminum support frame and offer sockets for the light sensors. This simplifies the adjustment of the cones on the sensors since all sensors are mounted on a single PCB (see Fig.~\ref{fig:focalplane}, left).

An additional, \SI{1}{\mm} thick Schott UG\,11~\cite{UG11} UV-pass filter with a size of \SI{6x6}{\mm} can be mounted in front of the photo sensors to improve the signal-to-noise ratio above \SI{400}{\nm} for the detection of fluorescence light by blocking a significant amount of the night sky ambient light. All components perform well in the near UV. Their transmittance curves are shown in Fig.~\ref{fig:efficiencies}.

\subsubsection{Photo sensors}

As photo sensors, Silicon photo-multipliers (SiPMs) have been chosen, which are cell-structured, photo-sensitive semiconductors. 
Their successful application was demonstrated by the First \mbox{G-APD} Cherenkov Telescope (FACT,~\cite{FACT}) which is operating since October 2011. It has proven the stable operation~\cite{FACT2}, as well as a significant increase of the duty cycle due the resistivity to bright light~\cite{FACTMoon}. 

Since the electrical properties of SiPMs change with the applied voltage, the precision of the provided voltage directly transforms into the stability of these properties. In particular, the so-called {\em over-voltage} needs to be kept sufficiently stable to achieve the required gain stability. The over-voltage itself is the difference between the applied voltage and the so-called {\em breakdown voltage}. As the breakdown voltage of each sensor is temperature dependent, a regulation circuit is required. In general, their temperature coefficient depends only on the type of SiPM but not on the individual sensor. Therefore, the power supply provides the bias voltage implementing an open-loop voltage correction based on the measured temperature. 


For the active part of the camera, 61 SiPM-arrays of type Hamamatsu S10943-3580X were utilized which are based on the Hamamatsu \mbox{S12573-100C}~\cite{HamamatsuS10943}. The S10943-3580X was an early prototype and features a special UV-transparent coating, with a peak photo detection efficiency of \SI{35}{\percent} at \SI{440}{\nm} (see Fig.~\ref{fig:efficiencies}, left). The next-generation SiPMs of type Hamamatsu S13360~\cite{HamamatsuS13360} or SensL MicroFJ-60035~\cite{SenslJ} feature a significant further increase of photo detection efficiency especially in the near UV and a remarkable decrease of correlated noise, such as dark counts and optical crosstalk. In particular, the enhanced photo detection efficiency improves the performance of the telescope.
 
Each SiPM-array is a \num{2x2}~matrix of single SiPM elements, each \SI{3x3}{\mm} in size with a cell-pitch of \SI{100}{\um}. All four elements share the same cathode but maintain individual anodes. All elements of one SiPM-array are read-out together connected in parallel. This yields a total sensitive area of \SI{6x6}{\mm} for every channel, each one coupled to one of the 61 Winston cones. 

\subsubsection{Focal plane}

All SiPM-arrays are mounted on a single printed circuit board (PCB) as shown in Fig.~\ref{fig:focalplane} (right). In total, the PCB holds 64 photo sensors. This includes three so-called \textit{blind pixels}. These three channels are shielded from the night-sky and they are used for monitoring of the electronic noise and SiPM gain during data-taking. They are also excluded from any trigger decision. 

The SiPM-to-SiPM distance on the equilateral triangular grid is 15.00\,mm. The PCB is fixed to the aluminum support frame with eight metric machine screws. Since aluminum is electrically conductive, no electronic components were mounted on the top side of the PCB except for the 64~SiPM packages and 64 temperature sensors of type LMT\,87 by Texas Instruments \cite{LMT87} directly underneath the sensors. The top side of the PCB is completely coated with solder mask and additionally isolated by a protective resin. 

The bottom side of the PCB hosts passive SiPM bias voltage filters and SiPM signal de-coupling. Surface mounted connectors for the SiPM bias voltages and signal output, as well as standard 2.54\,mm headers for the supply and analog output of the temperature sensors are mounted on the bottom side of the PCB.

Every channel profits from passive low-resistive low-pass bias voltage filtering of the 64 SiPM bias supplies. This reduces electronic noise in the high-frequency band above approximately\ \SI{30}{\kHz}. A low-resistive RC low-pass filter design with \SI{10}{\ohm} and \SI{470}{nF} was chosen to minimize the voltage drop introduced by the constant current drawn by the SiPMs during operation with a high background light level of the night sky. A small voltage-correction is still required under extreme conditions.
This amounts to a maximum of only \SI{1}{\percent} gain change at a \SI{1.4}{\volt} over-voltage over the full \SI{1.5}{\mA} load range. In addition, every channel is de-coupled with \SI{50}{\ohm} resistors and \SI{100}{nF} capacitors to match trace and coax cable wave-impedance. The voltage drop across the resistor leads to a maximum of 5\% change of the SiPM gain over the full load range. Operation during dark nights has shown that the typical current load is approximately\ \SI{100}{\uA} making a voltage correction for the combined loss optional. 

\section{Electronics}

To address the demand of a combined imaging air-Cherenkov and fluorescence telescope, the electronics were designed in a modular way. The SiPM signals are routed to two 80-pin headers on the focal-plane board. These connectors from the ERF\,8 series are manufactured by Samtec Inc.~\cite{SamtecErf8} and offer low signal insertion losses for frequencies up to 12\,GHz. A second board with mating connectors can be connected on top as an adapter connector. This PCB is application specific and groups the SiPM signals to board-to-wire connectors depending on the requested trigger layout and data acquisition system. This approach allows choosing from a variety of different data acquisition systems with different connectors whilst only re-designing the adapter board at reasonable costs.

\begin{figure}[thbp]
\centering
	\includegraphics[width=\textwidth]{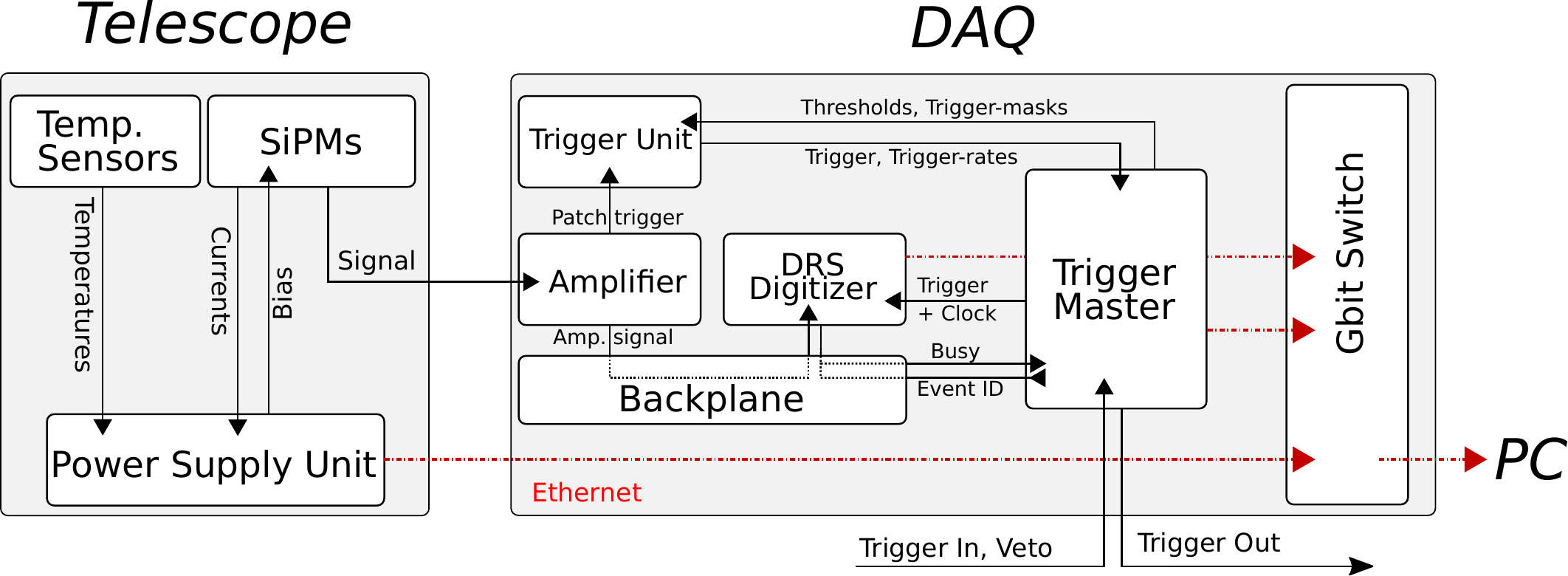}
\caption{\label{fig:DAQsys} The left box shows a schematic of the telescope interior including the SiPM based camera with temperature sensors and attached Power Supply Unit. The right box depicts the complete DAQ system including the Amplifier, Trigger Unit, Digitizer and the Trigger Master for trigger and clock distribution.}
\end{figure}

Figure \ref{fig:DAQsys} gives an overview over the whole system. In the following the main components of the electronics, data acquisition, trigger and power supply are described.

\subsection{Data acquisition}

\begin{figure}[thbp]
    \includegraphics[width=\textwidth]{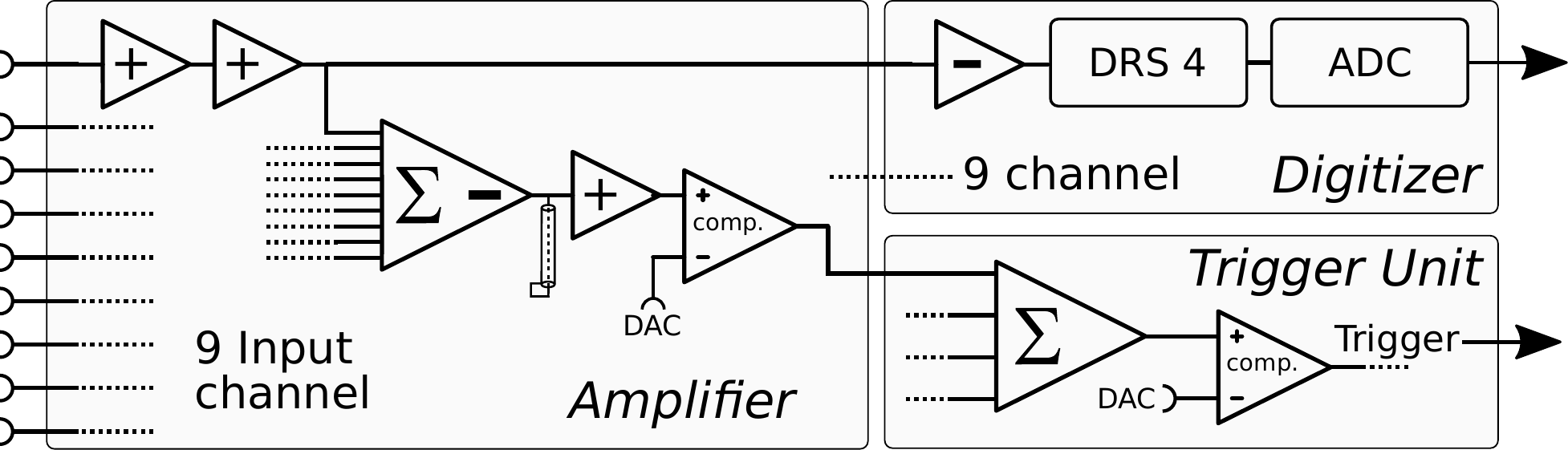}
	\caption{\label{fig:DAQsch} 
A simplified scheme of the signal path from the sensors to the data acquisition including the trigger logic.}
\end{figure}

\begin{figure}[thbp]
	\centering
    \includegraphics[width = .75\textwidth]{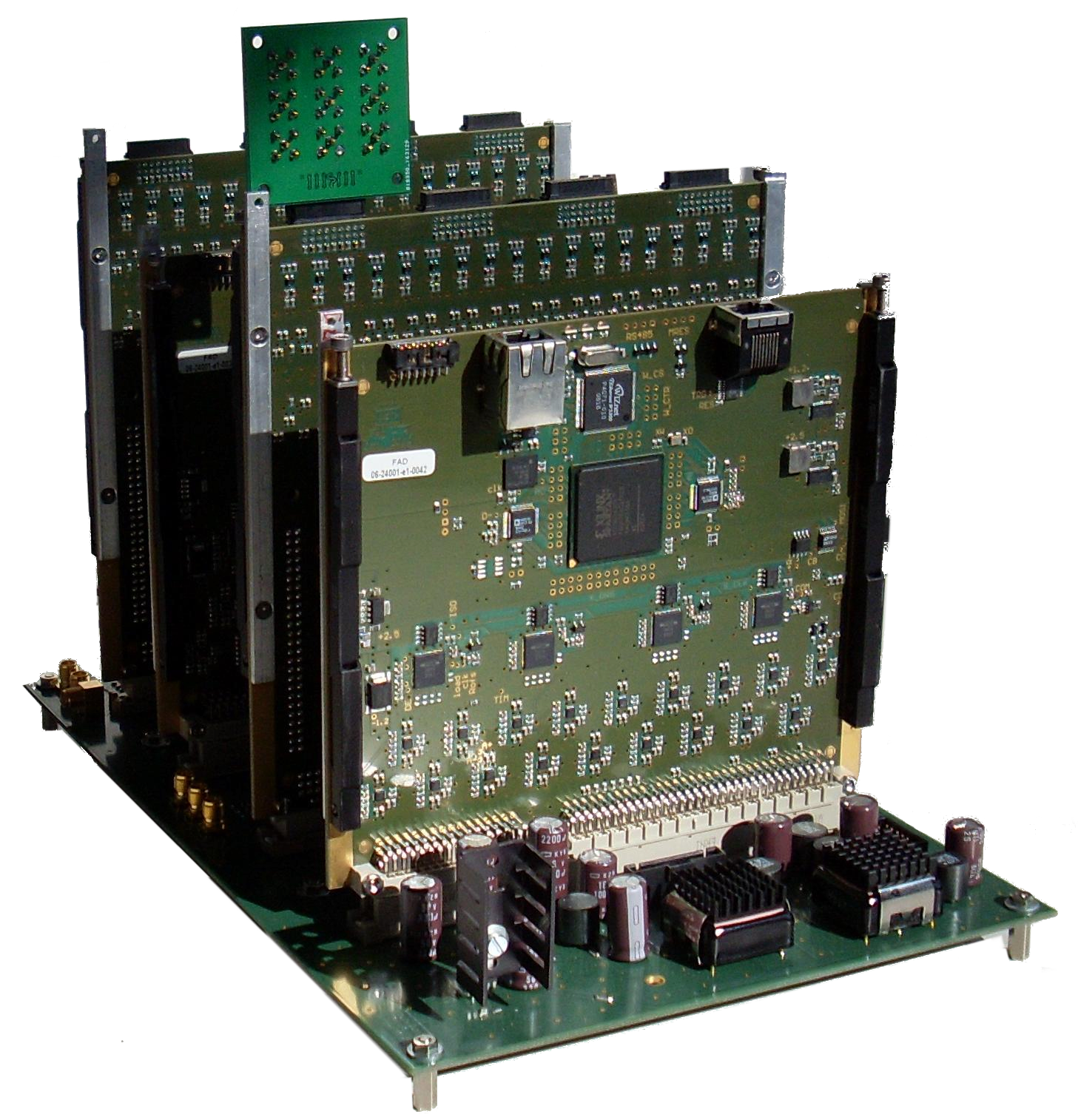}
\vspace{0cm}\caption{\label{fig:DAQ} 
Image of the data acquisition system. The backplane hosts two digitizer boards (first and third board) and two pre-amplifier boards (second and fourth board, trigger unit as mezzanine board on the back). In the front the integrated power supply on the backplane is visible.}
\end{figure}


To trigger and digitize the signals, the data acquisition system (DAQ) of the FACT project is utilized~\cite{FACT}. It has been developed for its \num{1440}-pixel SiPM-based camera and is designed modularly. Here, two 36-channel modules are sufficient to read out all of the 64 SiPM channels. One module consists of three different boards (see Fig.~\ref{fig:DAQ}). The first hosts the pre-amplifiers for 36 channels and generates analog sums of four groups of nine channels each. Each sum signal is clipped by a delay line shaping circuit to prevent pile-up. The shortened signal is discriminated using a comparator with a digital-to-analog converter (DAC) controlled threshold level (see Fig.~\ref{fig:DAQsch}).

The four discriminated signals are further processed by a mezzanine board, the trigger unit board, where the digitally programmable threshold voltages for the comparators are generated. A trigger counter for all four trigger channels and the output of the provided N-out-of-4 logic is realized in a field programmable gate array (FPGA). Communication takes place via serial RS\,485 connections. 

The pre-amplified analog signals are then passed to a data acquisition board. It hosts differential amplifiers to buffer the 36 signal channels. The buffered signals are routed to four 9-channel Domino Ring Sampling chips (DRS\,4,~\cite{DRS4}), a switched capacitor array to store the waveform in a ring buffer. The waveforms are digitized with commercial analog-to-digital converter (ADC) chips and further processed by an FPGA. Communication and data transmission takes place via a TCP/IP connection. Triggered events are tagged with an externally provided unique identifier via a RS\,485 bus.

Two complete modules are plugged in a back-plane, which routes the pre-amplified analog signals from the pre-amplifier to the data acquisition boards. A more detailed description of the readout electronics can be found in \cite{FACT}.

\subsection{Trigger}

\begin{figure}[thbp]
	\includegraphics[height=6.2cm]{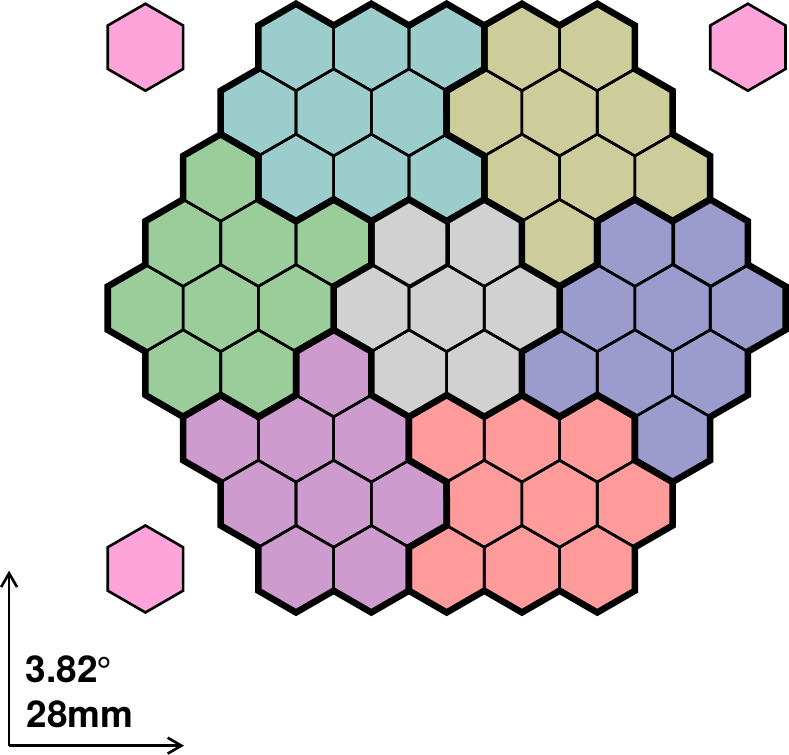}
	\hfill 
    \begin{tikzpicture}
	    \node[anchor=south west,inner sep=0] (image) at (0,0) {\includegraphics[height=6.0cm]{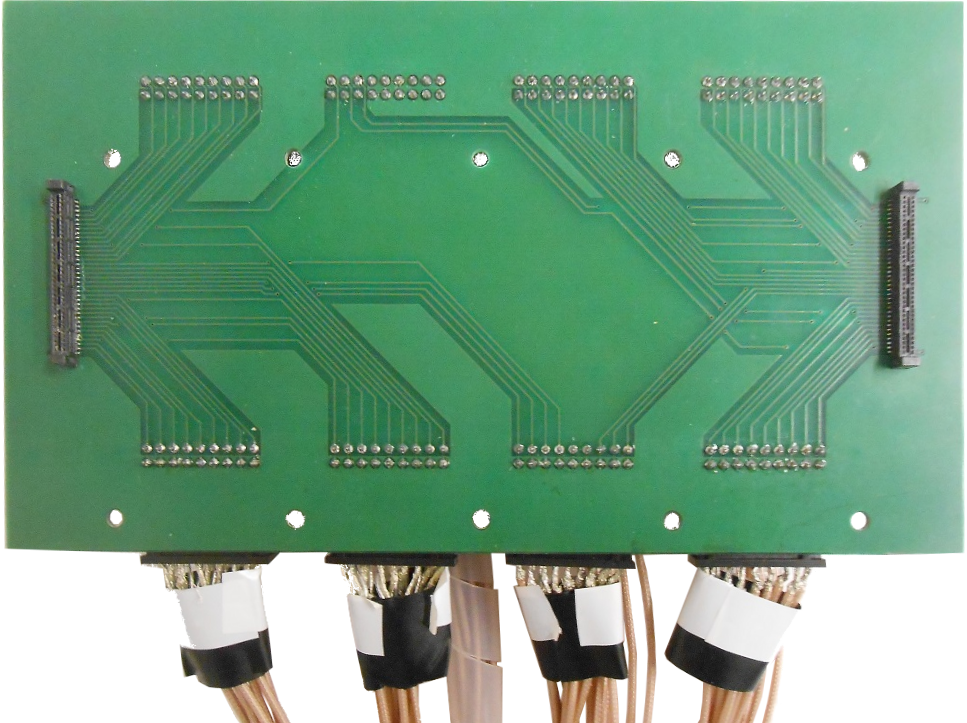}};
	    \begin{scope}[x={(image.south east)},y={(image.north west)}]
        	\draw[<->] (0.1-0.055, 0.1) -- (0.1+0.055, 0.1);
			\node at (0.1,0.1+0.05) {\SI{20}{\milli\meter}};
	    \end{scope}
	\end{tikzpicture}
\caption{\label{fig:trigger} \emph{Left:} Trigger layout of the telescope camera. The eight individual trigger groups of up to nine pixels each are color-coded. \emph{Right:} Adapter board used for the trigger routing.}
\end{figure}

Accounting for the provided trigger logic, the camera is virtually divided into eight trigger groups with up to nine pixels each. A layout of the trigger geometry is shown in Fig.~\ref{fig:trigger} (left). The grouping of the pixels is done on the adapter board as described above. The adapter board is shown in Fig.~\ref{fig:trigger}~(right).

The trigger layout has been optimized to minimize the bias caused by the geometry of air-shower images and provide a minimum complexity. The chosen layout distributes the compact nine-pixel trigger patches across the camera avoiding a preferential direction. The \textit{blind pixels} form a separate three-channel trigger patch which is not included in the main trigger decision.  

\subsection{Synchronization}

\begin{figure}[thbp]
	\includegraphics[height=5.7cm]{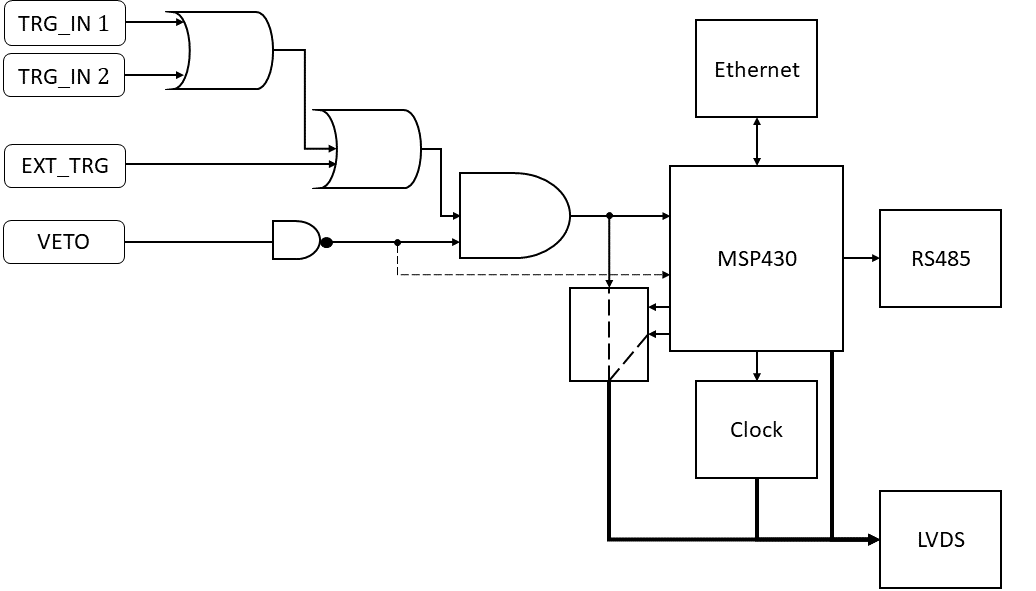}
	\hfill 
    \begin{tikzpicture}
	    \node[anchor=south west,inner sep=0] (image) at (0,0) {\includegraphics[height=5.9cm]{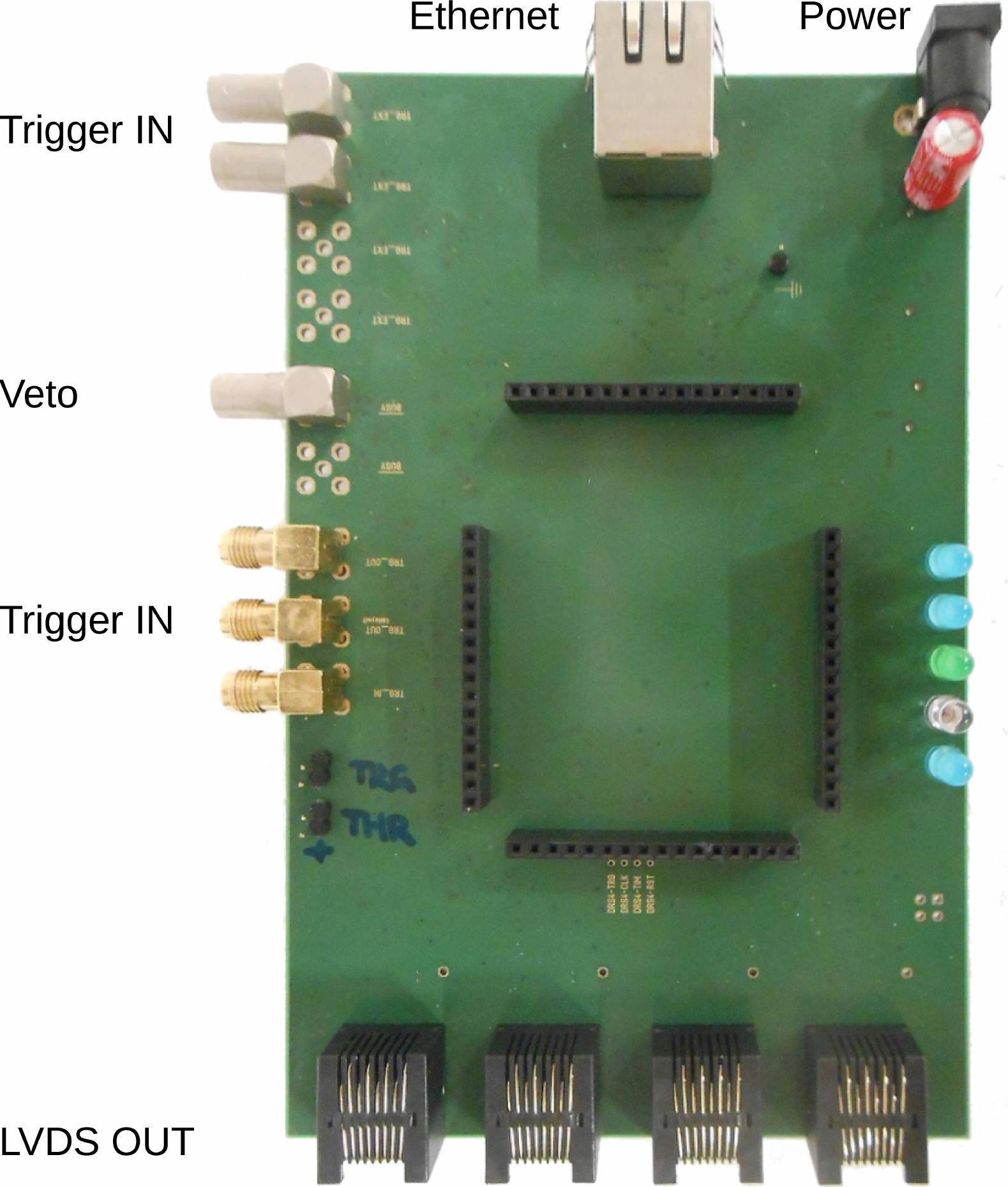}};
	    \begin{scope}[x={(image.south east)},y={(image.north west)}]
        	\draw[<->] (0.1-0.065, 0.1) -- (0.1+0.065, 0.1);
			\node at (0.1,0.1+0.05) {\SI{20}{\milli\meter}};
	    \end{scope}
	\end{tikzpicture}
\caption{\label{fig:miniftm} \emph{Left:} Schematic view of the trigger master. This electronic module handles the triggers from the FACT DAQ (LVDS) and sets the sampling rate for the DRS4 
chips (LVDS) and the event identifier tags (RS\,485). The module hosts a 
MSP\,430 
micro-controller and a Wiznet based Ethernet interface.
\emph{Right:} A picture of the Trigger Master board.
}
\end{figure}

Sampling and triggering of the two data acquisition modules need to be synchronized and the event tag has to be provided to them synchronously. For this purpose, the Trigger Master board has been designed (see Fig.~\ref{fig:miniftm}). It provides a synchronous clock to all DRS\,4 chips, implements the final trigger decision and provides the event tag via RS\,485. The board hosts a Texas Instruments MSP\,430 \cite{MSP430} micro-controller which is configurable via Ethernet implemented with a WIZnet W5500 \cite{W5500} Ethernet controller. It can be operated with any voltage between \SI{5}{\volt} and \SI{17}{\volt} and draws around \SI{70}{\mA} on \SI{5}{\volt}.

The clock for the sampling of the DRS\,4 \cite{DRS4} is provided by a programmable oscillator chip of type Linear Technology LTC\,6903, which is configured by the micro-controller. Although, the DRS\,4 allows for sampling rates from \SI{500}{MS\per\second} to \SI{5}{GS\per\second}, a sampling rate of \SI{2}{GS\per\second} is applied as default. The analog frequency value, which is a \nth{2048} of the sampling rate, is set via a corresponding Ethernet command.

The trigger condition is implemented with analog components reducing the time uncertainties due to the relatively slow micro-controller (\SI{32}{\MHz} clock max). The trigger output of both DAQ boards are combined by a logical OR including an optional external trigger. 

A veto logic is connected to the data acquisition boards where it is used to discard triggers whenever the trigger master itself or one of the data acquisition boards is busy, for example during data transmission.

A stop signal provided by the trigger master board stops writing to the ring buffer of the DRS\,4 after which the readout of a predefined number of cells starts. A readout window around the trigger position is achieved by delaying the stop signal by approximately\ \SI{300}{\ns} after the trigger. To avoid jitter, the delay is implemented as a fixed delay by chaining up two Texas Instruments SN74\,LVC1\,G123 chips. Given a sampling rate of \SI{2}{\GHz}, the full pipeline depth is around \SI{500}{\ns}. Thus, a delay of \SI{300}{\ns} shifts the trigger position to about one third of the readout trace of the full buffer is readout. The trigger is routed to the DAQ boards via LVDS lines (see Fig.~\ref{fig:miniftm}, left). 

To be able to process the triggers in the micro-controller, the analog pulse is stretched to more than 100\,ns guaranteeing that its interrupt catches the transition from low to high level. Whenever an interrupt occurs, the micro-controller sends a unique event identifier string via RS\,485 as broadcast to both data-acquisition boards. This unique identifier is provided in each event header to check for data consistency between two boards. The identifier contains an event counter which is later cross-checked with a provided internal event counter of each data acquisition board.

In addition, the trigger master board provides an independent RS\,485 communication with the trigger unit boards to set the trigger thresholds and read their trigger counters. This RS\,485 connection is implemented in the micro-controller via an independent TCP/IP socket separating commands for the controls of the data acquisition from the communication with the trigger units. 

\subsection{Power Supply Unit}

\begin{figure}[thbp]
	\includegraphics[height=6.3cm]{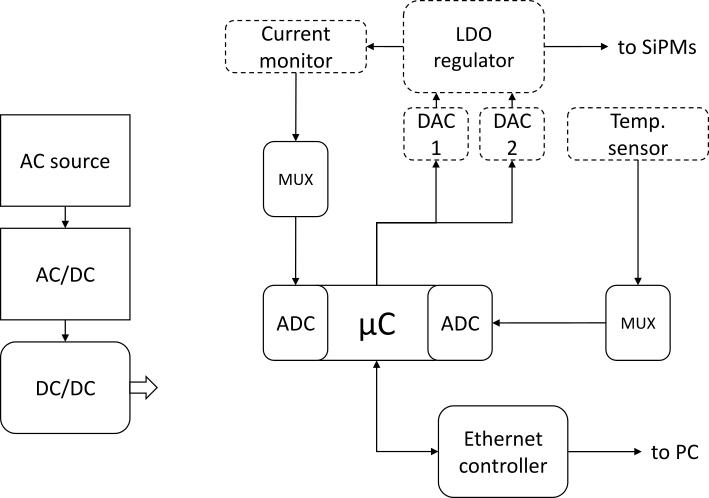}
	\hfill 
    \begin{tikzpicture}
	    \node[anchor=south west,inner sep=0] (image) at (0,0) {\includegraphics[height=5.9cm]{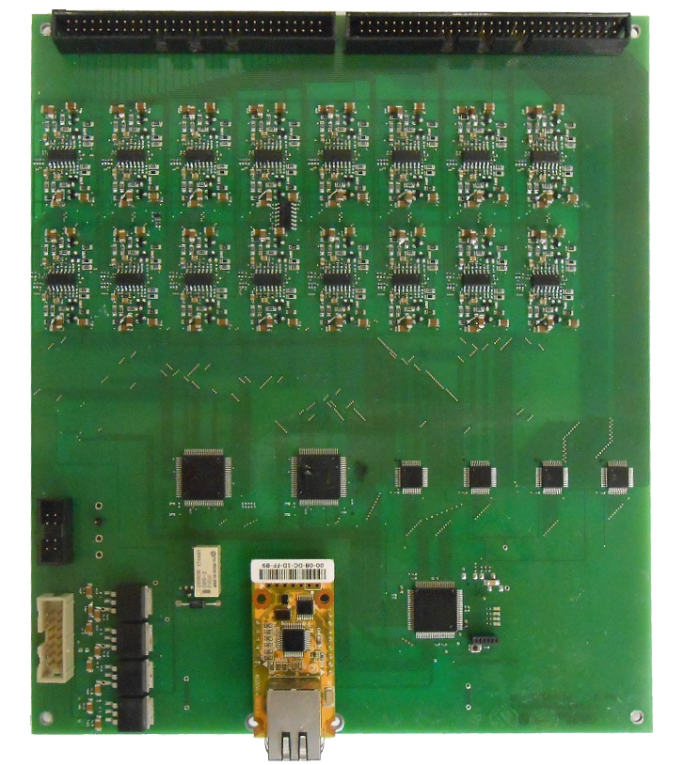}};
	    \begin{scope}[x={(image.south east)},y={(image.north west)}]
        	\draw[<->, color=white] (0.82-0.058, 0.2) -- (0.82+0.058, 0.2);
			\node[color=white] at (0.82,0.2+0.05) {\SI{25}{\milli\meter}};
	    \end{scope}
	\end{tikzpicture}
\caption{\label{fig:psu} Schematic view of the power supply \cite{PSU} (left) and a picture of the assembled bias power supply (right).}
\end{figure}

The Power Supply Unit is an autonomously working electronic system (see Fig.~\ref{fig:psu}, left). It is a proprietary design utilizing commercially available electronic components. It consists of two modules: The \emph{main module} provides the required supply voltages of the AC mains and the \emph{bias module} provides and regulates the bias voltage for the SiPMs.

The \emph{main module} is a transformer-based AC/DC converter unit which generates DC voltages from the AC lines of either \SI{110}{\volt} or \SI{230}{\volt} with \SI{50}{\Hz} or \SI{60}{\Hz}. This allows the system to operate with European and American mains. The transformers are followed by bridge-rectifiers and linear regulators. This design ensures that hardly any high-frequent noise is present on the voltage rails which is a requirement for the SiPM supplies. Two low voltage lines of +3.3\,V and +5\,V are used for the supply of digital and analog components of the bias supply module. The high voltage line of +73\,V is used by the bias module to generate the 64 individual SiPM bias voltages. The high voltage line is current-limited to approximately\ \SI{100}{\milli\ampere}. The main module also holds a highly efficient, yet low noise AC/DC switching converter which supplies both the data-acquisition electronics (limited to 5\,A on both $\pm$5\,V rails) and the trigger master board.

The \emph{bias module} holds 64 digitally-controlled linear regulators that are implemented with discrete components. Every channel can be programmed with a resolution of \SI{1}{\mV} individually on a voltage range between \SI{0}{\volt} and \SI{70}{\volt} \cite{PSU}. The maximal current that can be drawn by all channels is limited by the \SI{73}{\volt} supply of the main module to \SI{1.5}{\mA} per channel on average. The 64 external analog temperature sensors of type Texas Instruments LMT\,87, are connected to two 32-to-1 channel Multiplexer (MUX) chips of type Analog Devices ADG\,732. Their output nodes are connected to one of the on-chip 12-bit Analog-to-Digital Converter (ADC) ports of a Texas Instruments MSP\,430~\cite{MSP430} micro-controller. A Texas Instruments REF\,196 precision voltage chip is used for the analog reference of the ADC. The MSP\,430, a WIZnet W\,5500 \cite{W5500} Ethernet controller, eight 8-channel 16-bit Digital-to-Analog Converter (DAC) chips of type Analog Devices AD\,5668, two 32-channel 14-bit DAC chips of type Analog Devices AD\,5382, and two MUX chips of type Analog Devices ADG\,732 regulate and control the 64 linear regulators and the current measurement circuitry. The PCB of the second module is directly mounted behind the focal plane PCB eliminating the need of long, shielded cables.

Two application modes are implemented in the micro-controller: In the first mode, the power supply acts as a constant voltage source hardly dependent on load or temperature with \SI{0.4}{\mV\per\kelvin} stability~\cite{PSU}. The voltage for each channel can be programmed individually between \SI{0}{\volt} and \SI{70}{\volt} and the current and temperature for every channel can be read-out via Ethernet.

The second mode generates 64 individually programmable voltages with pre-determined linear dependence on the temperature. As the breakdown voltage of SiPMs is temperature-dependent (here approximately\ \SI{60}{\mV\per\kelvin}), the 64 analog temperature channels are read-out automatically and a linear progression of the bias voltage is adopted. Therefore, the power supply can be configured with a positive slope composed of an unsigned integer numerator and denominator in units of DAC-counts over ADC-counts (temperature sensors) and an offset given in units of DAC counts. The time interval for the automatic progression can be configured via Ethernet.

The current is determined at the \SI{73}{\volt} input of every regulator channel by measuring the differential voltage drop across a \SI{1}{\kilo\ohm} precision shunt resistor. The voltage drop at every channel is multiplexed and digitized by the 12-bit ADC of the MSP\,430. The current measurement resolution is approximately\ \SI{0.6}{\uA} (RMS) per bias channel. The value was improved over the one published in~\cite{PSU} of \SI{12}{\uA} (RMS) by the addition of \SI{\sim 1}{\Hz} low-pass filters in the current measurement circuitry. This is sufficient compared to the dark current of the SiPMs which is about \SI{1}{\uA} and compared to the night-sky background for dark nights which is typically \SI{100}{\uA}. This precise value is used to set the trigger threshold of the system. In addition, it allows to scan the I-V curves of every SiPM-pixel separately to calibrate the SiPMs in terms of breakdown voltage without the need of a DAQ. The maximum achievable Ethernet read-out rate of the power supply is approximately\ \SI{10}{\Hz}.

Every channel of the power supply system needs to be calibrated to assure that the absolute voltage and current scales are known and the resolution values from above apply to all channels.
The voltage calibration was done with a professional voltmeter. Here, two parameters sufficed to calibrate the physical voltage dependence on the digitally obtained value. The channel-to-channel variation of the calibration parameters are typically in the order of a few percent.
To calibrate the functional relation between the current of every channel and its digitized monitored value, a digitally controlled active load was developed. A simple linear function with two parameters was found to be sufficient to calibrate the current measurement.

\section{Data taking}

\subsection{System software}

The telescope is operated with software based on the FACT++ framework \cite{FACT} and has been extended with modules controlling the proprietary power supply and the Trigger Master. Debugging and testing has shown that all modules could be implemented easily.

The FACT software framework 
is designed as service-oriented architecture. Every hardware component is linked to a specific control program running on the host computer. To simplify maintenance, the framework implements a single program for each hardware component. The control programs are implemented as state machines, responding to changes of the hardware status or to commands submitted 
by the user or automatically through a central JavaScript interpreter. Inter-communication between individual programs is implemented with the Distributed Information Management system (DIM,~\cite{DIM}) developed at CERN. A Graphical User-Interface (GUI) allows to display quality-plots and events during operation. The FACT++ framework is described in detail in~\cite{FACT}.

\begin{figure}[thbp]
\centering
\includegraphics[width=\textwidth]{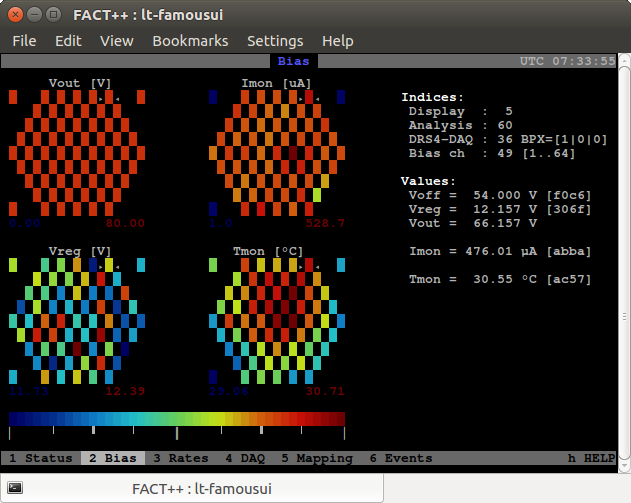} 
\caption{\label{fig:interface} An example screenshot of the light-weight console based display and remote access interface. Here the bias control page is visible showing the bias voltage (left camera views), the temperatures (bottom right view) and the high currents (upper right view) during a test on the roof top of RWTH Aachen University. The right column shows values related to the highlighted pixel which is selected with the cursor keys.}
\end{figure}

From the original FACT software, the control program for the data acquisition hardware~(\mbox{fadctrl}), the central data logger to store all logging output and slow control data (datalogger), the program to synchronize the actions of the different programs for individual runs (mcp) and a central JavaScript interpreter for scripting (dimserver) are used. 
New control programs have been implemented to control the trigger master board, communication with the trigger units to set the trigger threshold and read-out the trigger counters, and the control of the bias voltage supply. A dedicated program to calculate the trigger threshold as a function of the measured bias current for the particular system was implemented, as well as a console based user interface (see Fig.~\ref{fig:interface}) which allows to display several control parameters, steer major functions of the system, and a text-based camera display.

\FloatBarrier
\subsection{Start-up procedure}

\begin{figure}[thbp]
\centering
    \includegraphics[height=4.4cm]{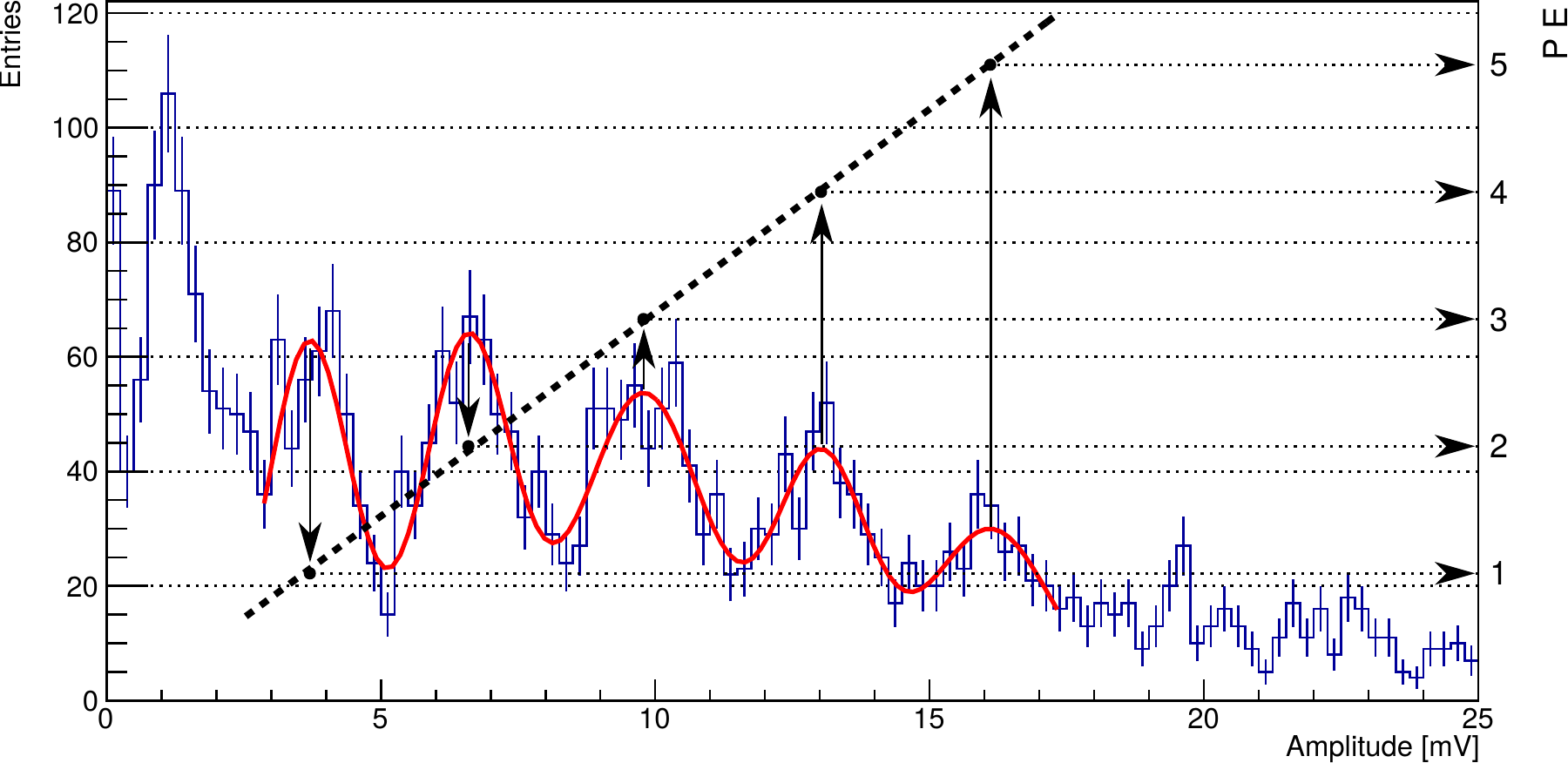}
    \hfill
	\includegraphics[height=4.4cm]{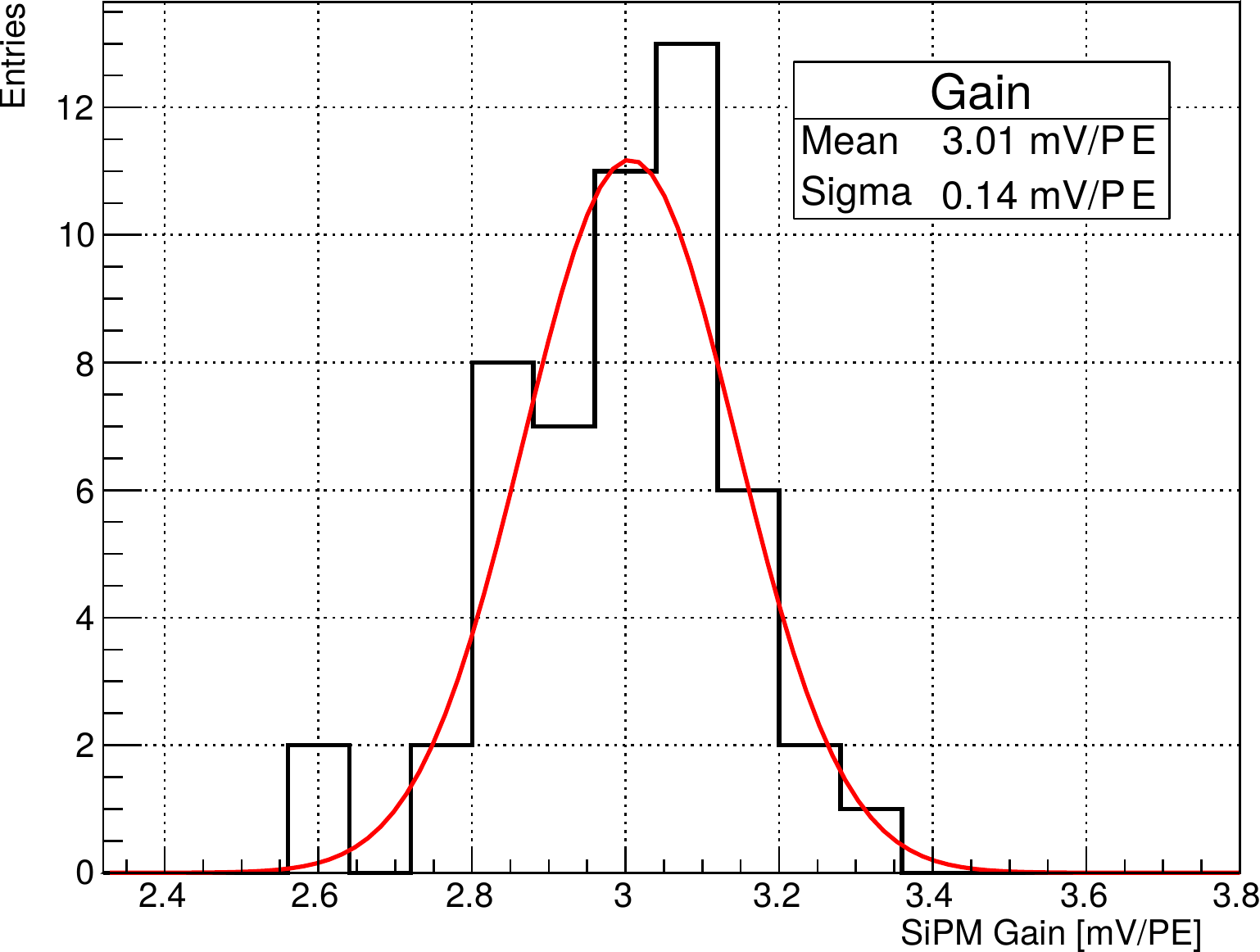}
\caption{\label{fig:fingerspec} \emph{Left:} In-situ measured dark-count spectra from a single SiPM and the gain determined by a simple global multi-Gaussian fit. The dark-count rate of the used sensor is \SIrange{8}{12}{MHz}~\cite{HamamatsuS10943}. \emph{Right:} Distribution of an exemplary gain calibration resulting in a mean gain of 3.01\,mV/PE with a sigma of 0.14\,mV/PE (4.7\,\%).}
\end{figure}

During the preparation for data taking, the hardware is configured according to the following procedure:
In a first step, the DRS\,4 baselines are calibrated. For this, the trigger master emits a trigger signal with a fixed rate of \SI{80}{\Hz}. After the calibration is done, the bias for the SiPMs can be set corresponding to the default over-voltage of \SI{1.4}{\volt}. 
When the lid of the telescope is closed, a SiPM gain calibration can be performed by using the fixed rate trigger of the trigger master to record DRS\,4 traces to form calibration histograms. The pulse height distribution of the dark-counts shows the excellent single photon resolution of the SiPMs. An exemplary result from a single pixel gain calibration can be seen in Fig.~\ref{fig:fingerspec} showing a single photo electron (PE) amplitude of $(3.12 \pm 0.09)$\,mV/PE.

\begin{figure}[thbp]
	\begin{center}
	\includegraphics[width=\textwidth]{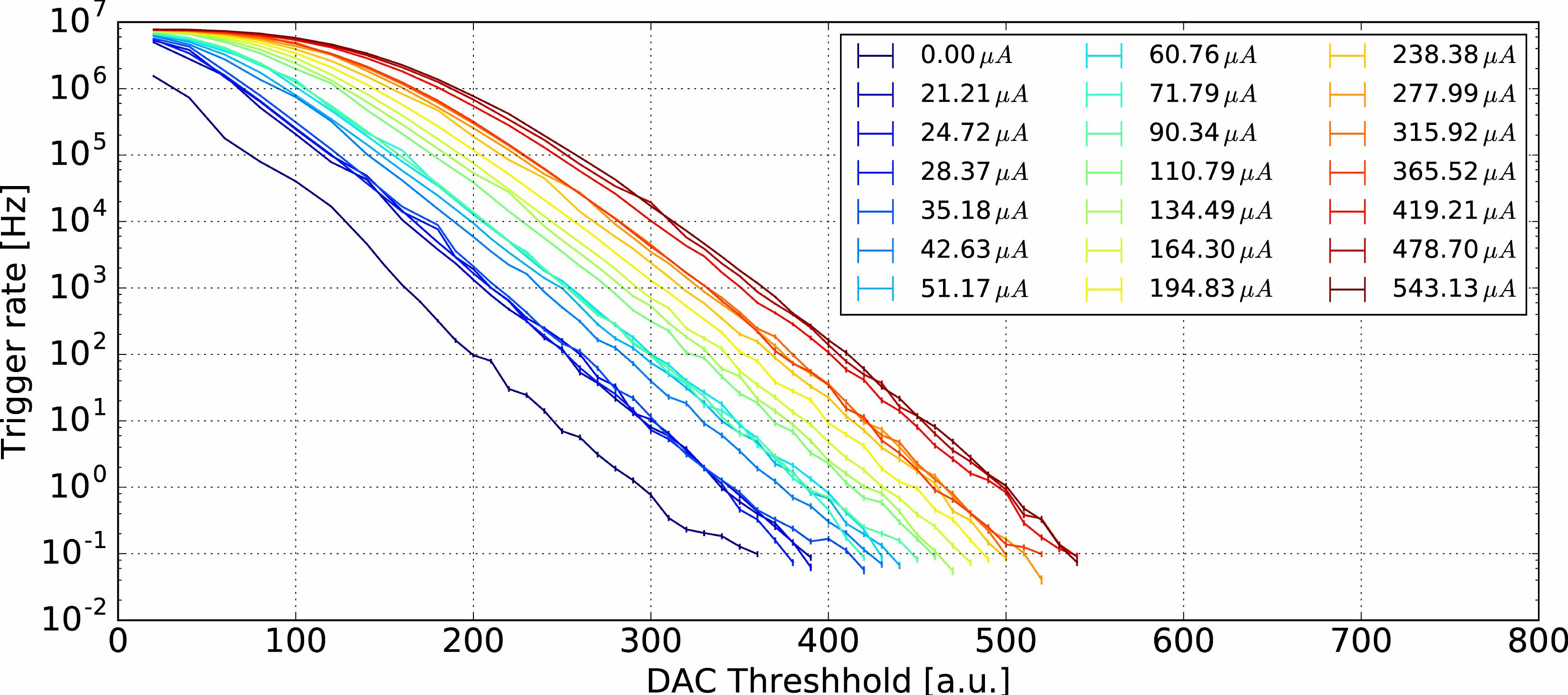}
	\end{center}
\caption{\label{fig:threscur} Trigger rate scans for different SiPM bias currents (incident light intensities). The dependency allows a direct estimation of the noise-rate by measuring the bias currents.  }
\end{figure}

With the lid open, the individual pixel current is obtained from the bias power supply and depending on the total current in the individual trigger patch, its trigger threshold is set such that the overall trigger rate corresponds to the value given in a configuration file, e.g.\ \SI{1}{\Hz}. The dependency of the noise trigger rate and the mean current of the trigger patches was calibrated using an isotropic light source in a distance of one focus length from the lens creating an almost homogeneous light distribution on the focal plane. Fig.~\ref{fig:threscur} show noise rate scans for different thresholds and bias currents (light intensities). Once the threshold is determined and applied, it is not changed during a single run. Changes of the night-sky background and the corresponding change in current of each pixel
requires to redo this procedure ~\cite{Biland:2014fqa}. The observations are split into single 5\,min runs with individual threshold settings. This automatic threshold adaption keeps the trigger rate stable, compensating for slow changes in the background light level e.g. from the moon or the sunrise. 

\begin{figure}[thbp]
	\includegraphics[width=\textwidth]{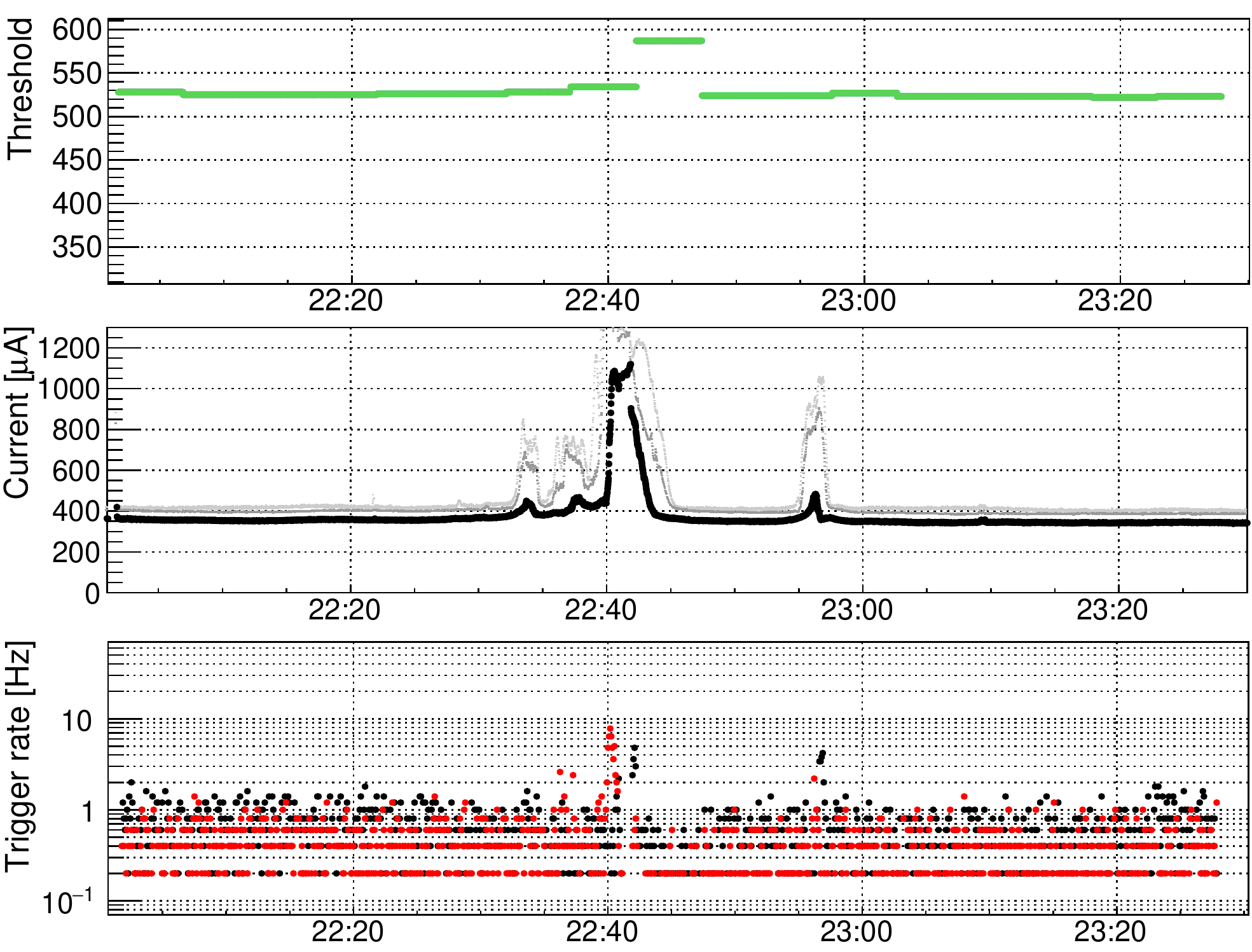}
\caption{\label{fig:qualityplot} Diagnostic plots taken during the night of the \nth{18} to \nth{19} of July 2017 in Aachen, Germany: Trigger threshold in arbitrary units (top), mean current per pixel in $\upmu$A (center, black) and of the brightest pixels (center, gray) and trigger rate in Hz (bottom). The conditions were stable over \SI{1.5}{\hour} except a short period of increased light from clouds. All times are UTC times.}
\end{figure}

A typical system log is given in figure~\ref{fig:qualityplot} showing the trigger rate, trigger thresholds and pixel currents measured during the night of the \nth{18} to \nth{19} of July 2017 in Aachen, Germany. Between 22:30 and 23:00 clouds reflected ambient light and strong transients in the bias current are observed. As the threshold adaption algorithm is not designed to compensate for fast or local increase of the background light level the trigger rate is also increased for a short period. Overall the system is stable and takes data  autonomously.

\section{Initial performance results}

\begin{figure}[thbp]
	\begin{center}
	\includegraphics[width=\textwidth, keepaspectratio]{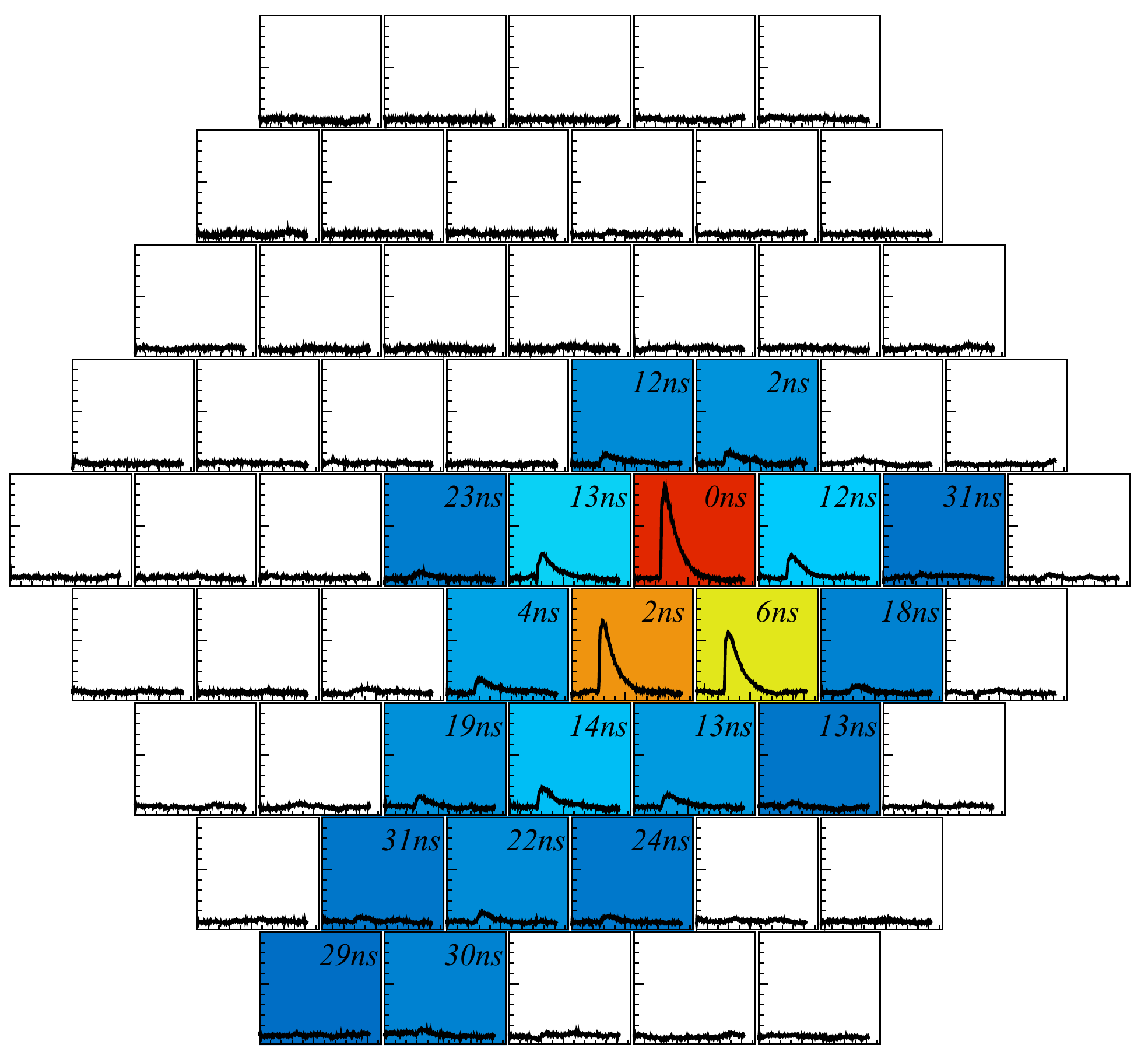}
	\end{center}
\caption{\label{fig:event} An air-shower event observed on June $19^{th}$ in Aachen, Germany. The timebase is \SIrange{0}{512}{\nano\second} and the amplitude range \SI{-80}{mV} to \SI{1000}{mV}. Pixels within a peak position of \SI{100}{ns} around the trigger position are color coded with respect to the individual signal amplitude. The delay of each maximum is given in respect to the earliest signal.}
\end{figure}

During the first commissioning of the telescope, several field test were carried out in Aachen, Germany (N\,\ang{50.780820}, E\,\ang{6.049149}) between June 2017 and July 2017. Due to the surrounding buildings and street lights, the additional stray light caused high bias currents of more than \SI{400}{\uA} per pixel. The utilized SiPM technology allowed a successful operation at a higher threshold. The telescope and the data acquisition system were placed on a trolley with the telescope vertically orientated (see Fig.~\ref{fig:telescope}, right). Many triggered events show the typical signature of air-shower events. A selected event is shown in Fig.~\ref{fig:event}. The event was recorded on June \nth{19} during a clear night with no moonlight. The arrival times (Fig.~\ref{fig:event}, right) show the expected coincident cluster around the trigger position at approximately\ \SI{300}{ns}. Pixels which do not contain a signal show randomly distributed arrival times.

In a similar configuration the telescope was successfully operated at the HAWC site at the Sierra Negra in Mexico at a altitude of \SI{4100}{\m} \cite{HAWCsEye}. The first results of the HAWC's Eye project will be published separately. For the IceAct project a seven pixel demonstrator was successfully operated at the South Pole during the antarctic winter 2016 and 2017 demonstrate the robustness of the design \cite{Auffenberg2017}. In 2018, a 61 pixel telescope was successfully commissioned at the South Pole. 

\section{Summary and Outlook}
The design of a general purpose compact telescope for the detection of Cherenkov or fluorescence light from air showers has been presented. This includes a detailed description of all major components including the enclosed Fresnel lens based optic, the 61 pixel SiPM camera, data acquisition and power supply systems. Results from the first light observed in an urban area showed a stable operation. An exemplary air shower image recorded during these field-tests was presented.

In table~\ref{cost}, a detailed breakdown of the cost for a single telescope and an estimate for 10+ telescopes is given. The majority of the cost is driven by the focal plane including the 61 SiPMs and their Winston Cones. The second cost factor is the required electronics. This does not yet include the data acquisition as the applied system was kindly provided by the FACT collaboration.

\newcommand{\ccb}{\cellcolor[gray]{0.85}}
\newcommand{\ccg}{\cellcolor[gray]{0.95}}
\begin{table}[htbp]
\centering
{\small
\begin{tabular}{||c|| r | r | l ||}
\hline
\ccb\textbf{Item} & \ccb\textbf{Price (1+)} & \ccb\textbf{Price (10+)} & \ccb\textbf{Comment}\\
\hline \hline
\ccg\textbf{Mechanics + Optics}	& \ccg\num{940} & \ccg\num{870}	  & \ccg\\
\hline
Carbon Fiber Tube			& \num{380}	  	   & \num{350}	      & e.g.\ Gade Kunststofftechnik GmbH\\
Fresnel Lens         		& \num{160}		   & \num{170} 	      & e.g.\ ORAFOL Fresnel Optics GmbH\\ 
Protective Glass Window     & \num{65}         & \num{50}         & e.g. Borofloat (Schott AG)\\
Plastic Support Structure   & \num{65}         & \num{50}         & Lens holder, camera holder\\
Diffusive Inlay             & \num{20}         & \num{20}         & \\
Camera Enclosure Box      	& \num{250}		   & \num{230}	      & Water tight\\
\hline\hline
\ccg\textbf{Focal plane} 	& \ccg\num{5,790}  & \ccg\num{3,930}  & \ccg\\
\hline
UG\,11 filter (\textit{optional}) &\num{640}		   & \num{500} 	      & e.g.\ Schott AG\\
Light Guides 				& \num{2,000}	   & \num{800}	      & Update: Solid cones (e.g.\ DentalChip GbR)\\
Photo Sensors: SiPM 		& \num{2,900}	   & \num{2,400}	  & e.g. SensL MicroFJ-60035-SMT\\
Focal Plane Board           & \num{250}        & \num{230}        & PCB + Components (w/o SiPM)\\
\hline\hline
\ccg\textbf{Electronics}	& \ccg\num{2,040}  & \ccg\num{1,400}  & \ccg\\
\hline
Data Acquisition (DAQ)		& ---        	   & ---     		  & FACT Loan (incl. pre-amp, trigger)\\
DAQ Backplane    			& \num{310}		   & \num{200}   	  & PCB + Components\\
Signal Routing Board        & \num{110}        & \num{60}         & PCB + Components\\
Signal Cable Connectors     & \num{210}        & \num{190}        & PCB + Components + Cables\\
Trigger Master Board        & \num{240}        & \num{120}        & PCB + Components\\
Bias Power Supply           & \num{920}        & \num{600}        & PCB + Components + Cables\\ 
Protective Housing          & \num{250}        & \num{230}        & \\
\hline\hline
\ccg\textbf{Others}			& \ccg\num{1,100}  & \ccg\num{900}    & \ccg\\
\hline
LV Power Supply + Cables    & \num{500}	       & \num{350}	      & Update (commercial: 80\,V, 12\,V, $\pm$5\,V)\\
PC + Storage + Network      & \num{500}        & \num{450}        & e.g. Intel NUC, 4\,TB HD, Switch, Cables\\
Pedestal                    & \num{100}        & \num{100}        & \\
\hline\hline
\ccb\textbf{Total}			& \ccb ca.~\num{10,000} & \ccb ca.~\num{7,500} & \ccb Excl.\ VAT\\
\hline
\end{tabular}
}
\caption{This table lists the estimated costs per telescope for quantities of one and ten telescopes in Euro (excl.\ VAT) as of December 2017. For components which are not commercially available anymore, comparable components are listed. For electronic boards, neither set-up costs, nor soldering nor assembly is included. Not listed are costs for consumables like screws, optical glue, etc.}
\label{cost}
\end{table}

Low costs and the enclosed design makes the telescope well suited for a variety of applications such as the extension of surface detectors and the deployment of huge telescope arrays. Its application in the projects IceAct and HAWC's Eye together with IceCube and HAWC, respectively, will be a benchmark for other future applications.

\paragraph{Outlook}

The presented design is the first revision of the telescope. A series of improvements are planned or already implemented to further improve the performance, durability and cost. Using new SiPM, like the SensL J-series, will significantly improve its detection efficiency. The implementation of solid light guides featuring a much higher geometrical fill factor will further improve the light collection efficiency. In total, an increase in effective aperture, and thus the corresponding decrease of the energy threshold by a factor four, is expected without a cost increase. The power supply unit was revised saving around \euro\,140 at the same performance. To save further on costs, the FACT data acquisition could be replaced by a solution utilizing the Target-C Application Specific Integrated Circuit (ASIC). The integrated module developed for the Cherenkov Telescope Array at costs of around \euro\,2400 includes 64-channels and a bias regulation circuit \cite{TargetC}. For the next revision of the telescopes, the application of commercial power supplies is foreseen. A more compact design will allow the integration of all components in a sealed enclosure attached to the telescope allowing much shorter signal cables. This all-in-one solution then only needs power and an Ethernet connection for data readout providing a convenient setup and operation.

\acknowledgments
This work has been supported by the Verbundforschung of the German Ministry for Education and Research (BMBF), funded by the Excellence Initiative of the German federal and state governments and the Helmholtz Alliance for Astroparticle Physics (HAP). We thank the German Academic Exchange Service
(DAAD) for their support and the FACT collaboration for support with the data acquisition and the Pierre Auger Observatory, as well as the HAWC and IceCube Collaborations for fruitful discussions during the development of this instrument. Furthermore, we thank Moritz Battermann for the measurement of the background light depended noise rates. Our special appreciation goes to the electronics workshops led by W.\ Feldhäuser and F.-P.\ Zantis, and the mechanics workshop led by D.\ Jahn and B.\ Philipps for their work on the telescope.

\newpage
\newcommand{\arXiv}[1]{{[\href{http://arxiv.org/abs/#1}{arXiv:#1}]}}
\newcommand{\TheHague}[1]{{[\href{https://pos.sissa.it/236/#1/pdf}{PoS(ICRC2015)#1}]}}
\newcommand{\Busan}[1]{{[\href{https://pos.sissa.it/301/#1/pdf}{PoS(ICRC2017)#1}]}}
\newcommand{\DOIarXiv}[2]{{[\href{http://dx.doi.org/#1}{doi:#1}, \href{http://arxiv.org/abs/#2}{arXiv:#2}]}}
\newcommand{\doi}[1]{{[\href{http://dx.doi.org/#1}{doi:#1}]}}

\newcommand{\ADS}[1]{{[\href{http://adsabs.harvard.edu/abs/#1}{ADS:#1}]}}



\begin{thebibliography}{99}

\bibitem{Kampert2012} 
	K.-H.~Kampert and A.~Watson, 
	``Extensive Air Showers and Ultra High-Energy Cosmic Rays: A Historical Review'',
  	Eur.\ Phys.\ J.\ H {\bf 37} (2012) 359
  	\arXiv{1207.4827}. 

\bibitem{Lorenz:2012nw} 
	E.~Lorenz and R.~Wagner, 
    ``Very-high energy gamma-ray astronomy: A 23-year success story in high-energy astroparticle physics'',
  	Eur.\ Phys.\ J.\ H {\bf 37} (2012) 459
  	\arXiv{1207.6003}.


\bibitem{Greisen1965} 
	K.~Greisen, 
    ``Highlights in air shower studies'', 
    In Proc. of the 9th International Cosmic Ray Conference (1965), Vol. 1, p.609 \ADS{1965ICRC....2..609G}.

\bibitem{Baltrusaitis:1985mx}
	R.~M.~Baltrusaitis et al. [Fly's Eye Collaboration],
	``The Utah Fly's Eye Detector'',
  	Nucl.\ Instrum.\ Meth.\ A {\bf 240} (1985) 410 \doi{10.1016/0168-9002(85)90658-8}.

\bibitem{AbuZayyad:2002sf}
	R.~U.~Abbasi et al. [HiRes Collaboration],
	``Monocular Measurement of the Spectrum of UHE Cosmic Rays by the FADC Detector of the HiRes Experiment'',
	Astropart.\ Phys.\  {\bf 23} (2005) 157
	\arXiv{astro-ph/0208301}.

  
\bibitem{Abraham:2009pm}
	J.~Abraham et al. [The Pierre Auger Collaboration],
	``The Fluorescence Detector of the Pierre Auger Observatory'',
	Nucl.\ Instrum.\ Meth.\ A {\bf 620} (2010) 227
	\arXiv{0907.4282}.

\bibitem{AbuZayyad:2012kk}
  	T.~Abu-Zayyad et al. [Telescope Array Collaboration], 
  	``The surface detector array of the Telescope Array experiment'',
  	Nucl.\ Instrum.\ Meth.\ A {\bf 689} (2013) 87
 	\arXiv{1201.4964}.
 
\bibitem{JelleyPorter1963} J. V. Jelley  and N. A. Porter. ``Cerenkov Radiation from the Night Sky.'' Quarterly Journal of the Royal Astronomical Society 4 (1963): 275 \ADS{1963QJRAS...4R.275J}.

\bibitem{Jackson:1998nia}
  	J.~D.~Jackson,
  	``Classical Electrodynamics'',
 	Wiley 1998, ISBN9780471309321.

\bibitem{Weekes:1989tc}
  T.~C.~Weekes et al.,
  ``Observation of TeV gamma rays from the Crab nebula using the atmospheric Cerenkov imaging technique'',
  Astrophys.\ J.\  {\bf 342} (1989) 379 \ADS{1989ApJ...342..379W}.

\bibitem{Aharonian:2006pe}
  	F.~Aharonian et al. [H.E.S.S. Collaboration],
  	``Observations of the Crab Nebula with H.E.S.S'',
  	Astron.\ Astrophys.\  {\bf 457} (2006) 899
  	\arXiv{astro-ph/0607333}.
 
\bibitem{Albert:2007xh}
  	J.~Albert et al. [MAGIC Collaboration],
  	``VHE Gamma-Ray Observation of the Crab Nebula and Pulsar with MAGIC'',
  	Astrophys.\ J.\  {\bf 674} (2008) 1037
  	\arXiv{0705.3244}.

\bibitem{Weekes:2001pd}
  	T.~C.~Weekes et al. [VERITAS Collaboration],
  	``VERITAS: The Very energetic radiation imaging telescope array system'',
 	Astropart.\ Phys.\  {\bf 17} (2002) 221
	\arXiv{astro-ph/0108478}.

\bibitem{TeVcat} TeVCat online source catalog [\url{http://tevcat.uchicago.edu}].


\bibitem{CTA}
  	M.~Actis et al. [CTA Consortium],
  	``Design concepts for the Cherenkov Telescope Array CTA: An advanced facility for ground-based high-energy gamma-ray astronomy'',
  	Exper.\ Astron.\  {\bf 32} (2011) 193
 	\arXiv{1008.3703}.
    
\bibitem{Niggemann2016} 
	T.~Niggemann, 
    ``The silicon photomultiplier telescope FAMOUS for the detection of fluorescence light,''
    PhD thesis, 
    RWTH Aachen University (2016),
    [\url{https://d-nb.info/1129876136/34}].
    
\bibitem{SenslJ} SensL J-Series, Datasheet (2017), [\url{http://sensl.com/downloads/ds/DS-MicroJseries.pdf}].

\bibitem{FACTHighlights} 
	D. Dorner et al. [FACT Collaboration], 
    ``FACT - Highlights from more than Five Years of Unbiased Monitoring at TeV Energies'',
    In Proc. of the \nth{35} ICRC \Busan{609}.
    
\bibitem{Haungs2015} 
	A.~Haungs et al. [The Pierre Auger Collaboration], 
    ``AugerNext: R\&D studies at the Pierre Auger Observatory for a next generation ground-based ultra-high energy cosmic ray experiment'',
    In Proc. of the \nth{34} ICRC 2015 \TheHague{593}.

\bibitem{Niggemann2013} 
	T.~Niggemann et al., 
    ``Status of the Silicon Photomultiplier Telescope FAMOUS for the Fluorescence Detection of UHECRs'',
    In Proc. of the \nth{33} ICRC (2013) \arXiv{1502.00792}.

\bibitem{Bretz2015} 
	T.~Bretz et al., 
    ``FAMOUS -- A fluorescence telescope using SiPMs''
    In Proc. of the \nth{34} ICRC \TheHague{649}.

\bibitem{PierreAuger} 
     A.~Aab et al., [The Pierre Auger Collaboration],
	``The Pierre Auger Cosmic Ray Observatory'',
	NIM A 798 (2015) 172-213 
	\DOIarXiv{10.1016/j.nima.2015.06.058}{1502.01323}.
    
\bibitem{HAWCsEye} 
	M. Schaufel, T. Bretz et al. [HAWC Collaboration], 
    ``Small size air-Cherenkov telescopes for ground detection arrays - a possible future extension?'', 
    In Proc. of the \nth{35} ICRC \Busan{786}.

\bibitem{HAWCCrab} 
	A. U. Abeysekara et al. [HAWC Collaboration], 
    ``Observation of the Crab Nebula with the HAWC Gamma-Ray Observatory'', 
    The Astrophysical Journal 843 (2017) 1
    \DOIarXiv{10.3847/1538-4357/aa7555}{1701.01778}.
    
\bibitem{Auffenberg2017} 
	J.~Auffenberg et al., 
    ``IceAct: Imaging Air Cherenkov Telescopes with SiPMs at the South Pole for IceCube-Gen2'',
    In Proc. of the \nth{35} ICRC \Busan{1055}.
 
\bibitem{Aartsen:2016nxy}
  M.~G.~Aartsen {\it et al.} [IceCube Collaboration],
  ``The IceCube Neutrino Observatory: Instrumentation and Online Systems'',
  JINST {\bf 12} (2017) P03012
  \DOIarXiv{10.1088/1748-0221/12/03/P03012}{1612.05093}.
  
\bibitem{Auffenberg2017b} 
	J.~Auffenberg et al., 
    ``On improving composition measurements by combining compact Cherenkov telescopes with ground based detectors'',
    In Proc. of the \nth{35} ICRC \Busan{404}.

\bibitem{Auffenberg2015} 
	J.~Auffenberg et al., 
    `` Design study of an air-Cherenkov telescope for harsh environments with efficient air-shower detection at 100 TeV'',
    In Proc. of the \nth{34} ICRC \TheHague{1155}.

\bibitem{PMMA} Orafol Fresnel Optics GmbH. “Private communications”. May 2011.

\bibitem{HamamatsuS10943} Hamamatsu S10943-series, Datasheets (2015).

\bibitem{UG11} Schott Advanced Optics GmbH. UV Bandpass UG11 (2017) [\url{http://www.schott.com/d/advanced_optics/2ac9d777-241a-4b09-940f-4dcf5f7aa898/1.2/schott-uv-bandpass-ug11-jun-2017-de.pdf}].  

\bibitem{Bouvier2013} 
	A. Bouvier et al.,
    ``Photosensor Characterization for the Cherenkov Telescope Array: Silicon Photomultiplier versus Multi-Anode Photomultiplier Tube'', 
    Hard X-Ray, Gamma-Ray, and Neutron Detector Physics XV. Vol. 8852. International Society for Optics and Photonics, 2013. \arXiv{1308.1390}.

\bibitem{Bunner1967} 
	A. N. Bunner, 
    ``Cosmic ray detection by atmospheric fluorescence'',
	PhD thesis, 
    February 1967,
    Cornell University [\url{http://nui118.jinr.ru/wiki/images/d/d9/Thesis_Buner.pdf}].
  
\bibitem{GEANT4} 
	J.~Allison et al.,
    ``Recent developments in Geant4'',
	Nuclear Instruments and Methods in Physics Research A 835 (2016) 186-225 \doi{10.1016/j.nima.2016.06.125}.  
	
\bibitem{Orafol} 
	Orafol Fresnel Optics GmbH. PositiveLinsen 1 (2017) 
[\url{http://www.orafol.com/tl_files/EnergyEurope/documents/PDF/Productlist/PositiveLinsen_1.pdf}].

\bibitem{EichlerMaster}
	H. M. Eichler,
    ``Characterisation studies on the optics of the prototype fluorescence telescope FAMOUS'',
    Masters thesis, 
    RWTH Aachen University, 2014 
    [\url{https://web.physik.rwth-aachen.de/~hebbeker/theses/eichler_master.pdf}].
    
\bibitem{OrafolOpt} Orafol Fresnel Optics GmbH. “Private communications”. June 2014.
  
\bibitem{Winston2005} 
	R.~Winston et al., 
    ``Nonimaging optics'', 
    Academic Press, 
    January 2005 \doi{10.1016/B978-0-12-759751-5.X5000-3}.  
  
\bibitem{JanPaulMaster}
	J.~P.~Koschinsky,
	``Entwicklung einer 61-Pixel Kamera für das IceAct Luftcherenkov-Teleskop am Südpol'',
	Masters Thesis, RWTH-Aachen University, 2017
	[\url{http://www.institut3b.physik.rwth-aachen.de/global/show_document.asp?id=aaaaaaaaaaxbjez}].

\bibitem{BretzRibordy} 
	T. Bretz and M. Ribordy, 
    ``Design constraints on Cherenkov telescopes with Davies-Cotton reflectors'', 
    Astroparticle Physics 45 (2013) 44--55 
    \DOIarXiv{10.1016/j.astropartphys.2013.03.004}{1301.6556}.

\bibitem{Rakic1998} 
	A. D. Rakic et al., 
    ``Optical properties of metallic films for vertical-cavity optoelectronic devices'',
    Appl. Opt. 37.22 (1998), pp. 5271–5283 \doi{10.1364/ao.37.005271}.

\bibitem{FACT} 
	H.~Anderhub et al. [FACT Collaboration],
    ``Design and Operation of FACT -- The First G-APD Cherenkov Telescope'',
    JINST 8 (2013) P06008
    \doi{10.1088/1748-0221/8/06/P06008}.

\bibitem{FACT2} 
	A.~Biland et al. [FACT Collaboration],
    ``Calibration and performance of the photon sensor response of FACT — the first G-APD Cherenkov telescope'',
 	JINST 9 (2014) P10012 
    \doi{10.1088/1748-0221/9/10/P10012}.

\bibitem{FACTMoon} 
	M. L. Knoetig et al. [FACT Collaboration], 
    ``FACT - Long-term stability and observations during strong Moon light'', 
    In Proc. of the \nth{33} ICRC (2013) \arXiv{1307.6116}.
    
\bibitem{HamamatsuS13360} Hamamatsu S13360-series, Datasheets (2017), [\url{http://www.hamamatsu.com/resources/pdf/ssd/s13360_series_kapd1052e.pdf}].

\bibitem{LMT87} Texas Instruments LMT87 analog temperature sensor datasheet, [\url{http://www.ti.com/lit/ds/symlink/lmt87.pdf}].

\bibitem{SamtecErf8} Samtec Inc, ERF8 / ERM8 datasheet,  [\url{http://suddendocs.samtec.com/catalog_english/erf8.pdf}].

\bibitem{DRS4} 
	Stefan Ritt et al., 
    ``Application of the DRS chip for fast waveform digitizing, Technology and instrumentation in particle physics'', 
    In Proc. 1st International Conference (TIPP09), Tsukuba, Japan, March 12-17, 2009 \arXiv{1407.8059}.




\bibitem{MSP430} 
	Texas Instruments Inc., 
    ``MSP\,430 ultra-low-power microcontrollers'' 
    [\url{http://www.ti.com/Mikrocontroller/MSP}].

\bibitem{W5500} 
	WIZnet Co. Ltd., 
    ``W5500 hardwired TCP/IP embedded Ethernet controller'' 
    [\url{http://www.wiznet.io/product-item/w5500}].

\bibitem{PSU} 
	J.~Schumacher et al., 
    ``Dedicated power supply system for silicon photomultipliers '',
    In Proc. of the \nth{34} ICRC \TheHague{605}.

\bibitem{DIM} 
	Distributed Information Management System (DIM) 
    [\url{http://dim.web.cern.ch}].

\bibitem{Biland:2014fqa}
  A.~Biland et al. [FACT Collaboration],
  ``Calibration and performance of the photon sensor response of FACT -- The First G-APD Cherenkov telescope'',
  JINST {\bf 9} (2014) 10 
  \arXiv{1403.5747}.

\bibitem{TargetC} S.~Funk et al. [CTA Consortium],
  ``TARGET: A digitizing and trigger ASIC for the Cherenkov telescope array'',
  AIP Conf.\ Proc.\  {\bf 1792} (2017) 080012 \arXiv{1610.01536}.


\end{thebibliography}
\end{document}